\documentclass[prb,aps,amsmath,preprint]{revtex4}
\usepackage{graphicx}
\usepackage{subfigure}
\usepackage{epsfig}
\usepackage{dcolumn}

\usepackage{bm}
\usepackage{amsmath}
\usepackage[dvips]{color}

\begin{document}
\title{Density functional study of electronic structure, elastic and optical properties of MNH$_2$ (M=Li, Na, K, Rb)}

\author{K. Ramesh Babu
  and G. Vaitheeswaran*}
\affiliation {Advanced Centre of Research in High Energy Materials (ACRHEM),
University of Hyderabad, Prof. C. R. Rao Road, Gachibowli, Andhra Pradesh, Hyderabad - 500 046, India
}
\date{30 May 2014}
\begin{abstract}
 We report systematic first principles density functional study on the electronic structure, elastic and optical properties of nitrogen based solid hydrogen storage materials LiNH$_2$, NaNH$_2$, KNH$_2$, and RbNH$_2$. The ground state structural properties are calculated by using standard density functional theory and also dispersion corrected density functional theory. We find that van der Waals interactions are dominant in LiNH$_2$ whereas they are relatively weak in other alkali metal amides. The calculated elastic constants show that all the compounds are mechanically stable and LiNH$_2$ is found to be stiffer material among the alkali metal amides. The melting temperatures are calculated and which follows the order RbNH$_2$ $<$ KNH$_2$ $<$ NaNH$_2$ $<$ LiNH$_2$. The electronic band structure is calculated by using the Tran-Blaha modified Becke-Johnson potential and found that all the compounds are insulators with a considerable band gap. The [NH$_2$]$^-$ derived states are completely dominating in the entire valence band region while the metal atom states occupy the conduction band. The calculated band structure is used to analyze the different interband optical transitions occur between valence and conduction bands. Our calculations show that these materials have considerable optical anisotropy.
\end{abstract}
\maketitle
\section {Introduction}
 Hydrogen, being the most abundant element in the universe is considered to be the best choice as future energy source. The superior qualities of hydrogen such as light weight, high energy per unit mass and eco-friendly combustion products enabled it a suitable replacement for current carbon based energy resources. However storage of hydrogen is a difficult problem. To store hydrogen as a compressed gas, high pressure tanks are required and the storage in liquid state needs insulated cryogenic tanks. Both these are not suitable for practical applications as storage in gaseous state leads to low energy density while safety and cost are major concerns for the storage in liquid state. Therefore solid state hydrogen storage is the only choice to look at the issue and it requires search for solid materials that can store hydrogen with volumetric densities greater than those of other states of hydrogen storage. \cite{Guo, Vinod, Orimo, Jain, Umegaki}
\paragraph*{}
 Hydrides of light and heavy metals exhibit qualitative solid-state hydrogen storage properties such as high volumetric densities and low pressure desorption. \cite{Vinod} But, the slow absorption and desorption kinetics limits their practical use. On the other hand, complex hydrides consisting of light metal elements have been noticed as materials with fast hydrogen kinetics and most importantly they can store hydrogen in greater percentages over simple metal hydrides. Complex metal hydrides with potential solid state hydrogen storage applications are classified as alanates, borohydrides, and amides. \cite{Orimo} Among the complex metal hydrides, nitrogen based materials namely amides received much interest because of their high hydrogen storage capacity and low operating temperatures. \cite{Umegaki} Most importantly, alkali metal amides gained interest because of their potential reversible hydrogen storage applications. \cite{Chen}
\paragraph*{}
The first report on alkali metal amides KNH$_2$ and NaNH$_2$ appeared in the beginning of 19th century by the authors Gay Lussac and Thenard and later in 1894, Titherley reported the synthesis of LiNH$_2$. \cite{GayLussac, Titherley} These metal amides were traditionally used as reagents in organic synthesis. \cite{Fieser, Mulvey} However, it is only in the year 2002, Chen et al. discovered LiNH$_2$ as a potential candidate for reversible hydrogen storage applications and hence paved new insights in to the hydrogen storage application of these materials. \cite{Chen} In general for any solid state hydrogen storage system, the important characteristics are the following: the material should have high volumetric as well as gravimetric hydrogen densities, fast hydriding and dehydriding characteristics, and suitable thermodynamic properties. It is a known fundamental aspect that the hydrogen absorption and desorption properties can be well judged through the knowledge of electronic structure and bonding. Moreover, the physical properties such as the elastic and optical properties are much important to understand the mechanical stability and optical response of solid state hydrogen storage materials. In addition, these are necessary in determining the thermodynamic properties of these kind of materials. \cite{Li}
\paragraph*{}
Theoretical calculations based on density functional theory are accurate enough to predict and reproduce the experimentally measured quantities. The electronic structure of LiNH$_2$ was reported by using first principles plane wave pseudopotential calculations and the authors found that the material has non-metallic nature. \cite{Miwa, Herbst} Recently, the electronic structure of LiNH$_2$ was determined through XAS studies and the results are found to compliment the earlier theoretical reports. \cite{Kamakura} Elastic constants of LiNH$_2$ were reported theoretically by using the density functional theory calculations. \cite{Hector} The high pressure behaviour of LiNH$_2$ was reported by using the Raman spectroscopy technique and found that the system undergo a phase transition from ambient tetragonal $\alpha$-LiNH$_2$ structure to high pressure $\beta$-LiNH$_2$ phase. \cite{Raja} Theoretically pressure-induced structural phase transitions of LiNH$_2$ was reported by using the ab-initio total energy calculations and evolutionary structure prediction simulations. \cite{Rao, Prasad} By using combined synchrotron X-ray diffraction measurements and ab initio density functionals, the pressure-induced phase transition with large volume collapse (11$\%$) in LiNH$_2$  was reported. \cite{Zou} The high pressure behaviour of NaNH$_2$ was reported by using Raman and Infrared spectroscopies \cite{Song} and also through evolutionary structure simulations. \cite{Oganov} To the best of our knowledge, there are no comparative studies available to explain the structural, electronic, elastic and optical properties of the alkali metal amides LiNH$_2$, NaNH$_2$, KNH$_2$ and RbNH$_2$. Hence, in this present work we aim to study these properties  by performing density functional theory calculations. The rest of the paper is organized as follows: section 2 deals with the computational details and theoretical methods used in the present study. Results and discussion are presented in section 3. Finally in section 4, we end with a brief conclusion of the present study.
 \section {Computational  details}
The density functional calculations are carried out with the plane wave pseudopotential method \cite{Segall}. Vanderbilt type ultrasoft pseudopotentials are used for the present calculations. \cite{Vanderbilt} The electronic wave functions are obtained by using density mixing scheme and the structures are relaxed using the Broyden, Fletcher, Goldfarb, and Shannon (BFGS) method. The local density approximation with Ceperley-Alder\cite{Ceperley} exchange-correlation potential parameterized by Perdew and Zunger \cite{PPerdew} (LDA-(CA-PZ)), and also generalized gradient approximation proposed by Perdew-Burke-Ernzerhof \cite{Perdew}(GGA-PBE) have been used to describe the exchange-correlation potential. The pseudo atomic calculations were performed for Li 2s$^1$, Na 2s$^2$ 2p$^6$ 3s$^1$, K 3s$^2$ 3p$^6$ 4s$^1$, Rb 4s$^2$ 4p$^6$ 5s$^1$, N 2s$^2$ 2p$^3$ and H 1s$^1$, respectively. The cut-off energy for plane waves is set to 380 eV for LiNH$_2$, 400 eV for  NaNH$_2$ and 320 eV for KNH$_2$, RbNH$_2$, respectively. Brillouin zone sampling is performed by using the Monkhorst-Pack scheme \cite{Monkhorst} with a k-point grid of 4x4x2 for LiNH$_2$, 4x3x4 for NaNH$_2$ and 4x4x4 for KNH$_2$, RbNH$_2$, respectively. The values of both plane wave cut-off energy and k-point grid are determined by performing total energy calculations to achieve the convergence of 1meV. For the computation of electronic band gap, we use the recently developed Tran Blaha-modified Becke Johnson potential \cite{TB} within linearized augmented plane wave (LAPW) method as implemented in WIEN2k package. \cite{DJ, PBlaha}
\paragraph*{}
It is well known that the standard density functional methods are not adequate enough to treat the weak dispersive interactions present between the molecules of the crystal. To treat vdW interactions efficiently, recently Grimme successfully applied the vdW correction to the exchange - correlation functional of standard density functional theory (GGA-PBE+G06) at semi empirical level. \cite{Grimme} According to semi-empirical dispersion correction approach proposed by Grimme, the total energy of the system can be expressed as
\begin{equation}
E_{total} = E_{DFT} + E_{Disp}
\end{equation}
where
\begin{equation}
E_{Disp} = s_i\Sigma_{i=1}^N\Sigma_{j>i}^Nf(S_RR^{0}_{ij}, R_{ij})C_{6, ij}R_{ij}^{-6}
\end{equation}
here C$_{6, ij}$ is called dispersion coefficient between any atom pair $i$ and $j$ which solely depends upon the material and R$_{ij}$ is the distance between the atoms $i$ and $j$ respectively. In the present calculations, the C$_{6, ij}$ values of Li, Na, K and Rb are 1.61 J-nm$^6$/mol$^{-1}$, 5.71 J-nm$^6$/mol$^{-1}$, 10.80 J-nm$^6$/mol$^{-1}$ and 24.67 J-nm$^6$/mol$^{-1}$ and the corresponding R$^0$$_{ij}$ values of 0.825 \AA for Li, 1.144 \AA for Na, 1.485 \AA for K and 1.628 \AA for Rb have been used. Whereas for the case of N atom we have used C$_{6, ij}$ value of 1.23 J-nm$^6$/mol$^{-1}$ and R$^0$ $_{ij}$ value of 1.397 \AA and for H atom the corresponding values are 0.14 J-nm$^6$/mol$^{-1}$ and 1.001$\AA$ respectively. Starting from the optimized crystal structure all the related properties have been calculated. The elastic constants were calculated by using stress-strain method as implemented in the CASTEP code.

 \section{Results and discussions}
\subsection{Structural properties}
At ambient conditions, LiNH$_2$ crystallizes in tetragonal structure with space group I-4 (82) (z=4) (shown in Figure 1a) \cite{Jacobs} and NaNH$_2$ crystallize in orthorhombic structure with space group Fddd (70) (z=16)  (shown in Figure 1b). \cite{Nagib} Whereas both KNH$_2$ and RbNH$_2$ crystallize in monoclinic structure with space group P2$_1$/m (13) (z=2)  (shown in Figure 1c). \cite{Keeffe} The corresponding reciprocal lattices with Brillouin zone for LiNH$_2$, NaNH$_2$, and KNH$_2$ (same for RbNH$_2$) are also shown in the Figure 2(a), 2(b) and 2(c) respectively. All the calculations are carried out by adopting the experimental crystal structures as the initial structures and they are relaxed to allow the ionic configurations, cell shape, and volume to change in order to predict the theoretical equilibrium crystal structure within standard density functionals LDA (CA-PZ), GGA (PBE) and with dispersion corrected density functional GGA (PBE+G06). The optimized lattice parameters and internal atomic positions of metal atom M (Li, Na, K, Rb, Cs), nitrogen N and hydrogen H are tabulated in Table I and Table II along with experimental data respectively. It can be seen that the calculated GGA (PBE) values are in good agreement with experiment when compared to LDA (CA-PZ) and GGA (PBE+G06) results. This implies that the dispersion interactions do not play prominent role in these systems as the volume computed with PBE+G06 functional results in large errors compared to PBE volume, except for the case of LiNH$_2$. Therefore, to calculate the elastic constants, electronic and optical properties we adopted the GGA (PBE) functional.

\begin{figure}
\centering
\subfigure[]{\includegraphics[height=5cm,width=6cm]{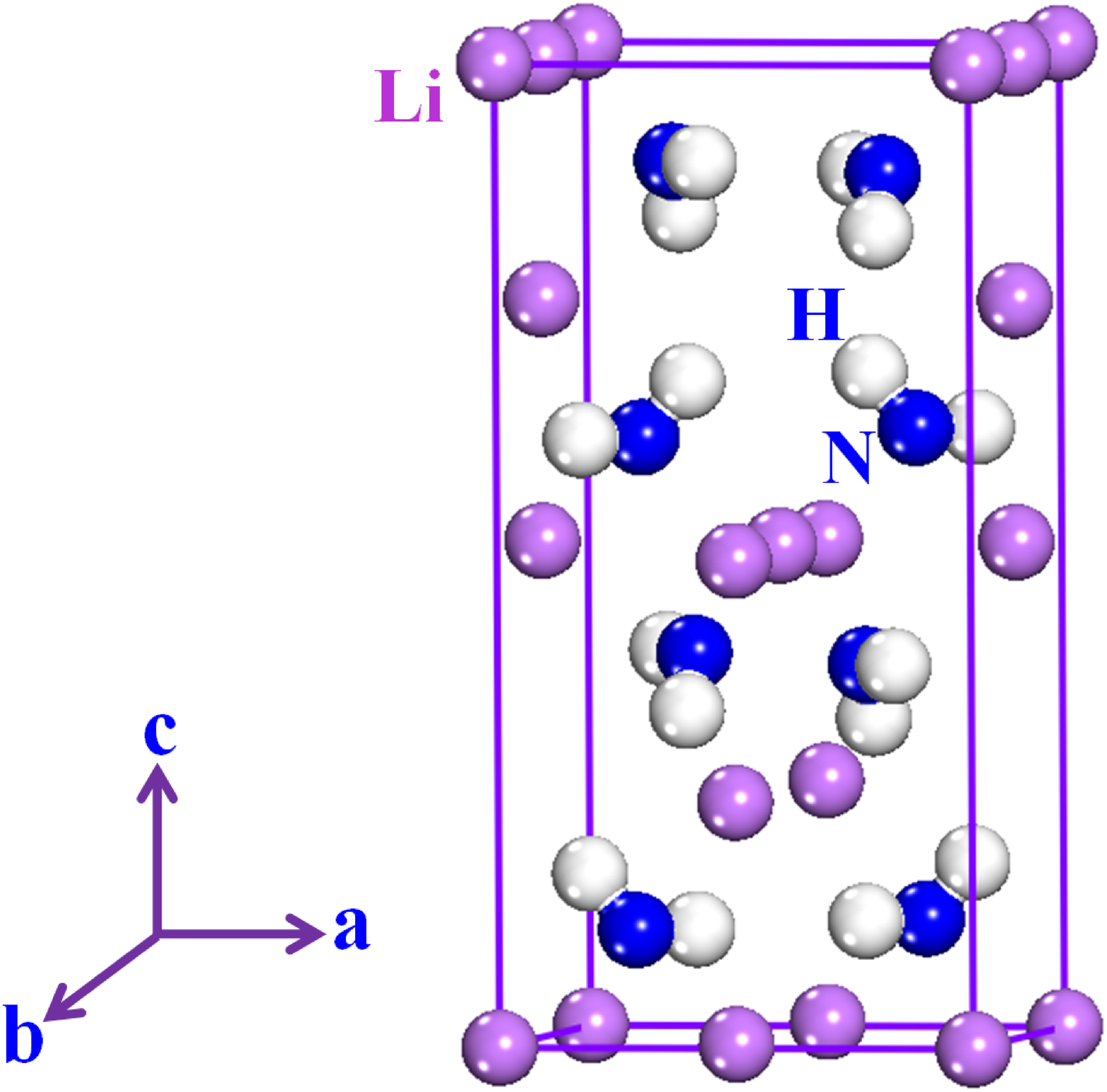}}
\subfigure[]{\includegraphics[height=5cm,width=6cm]{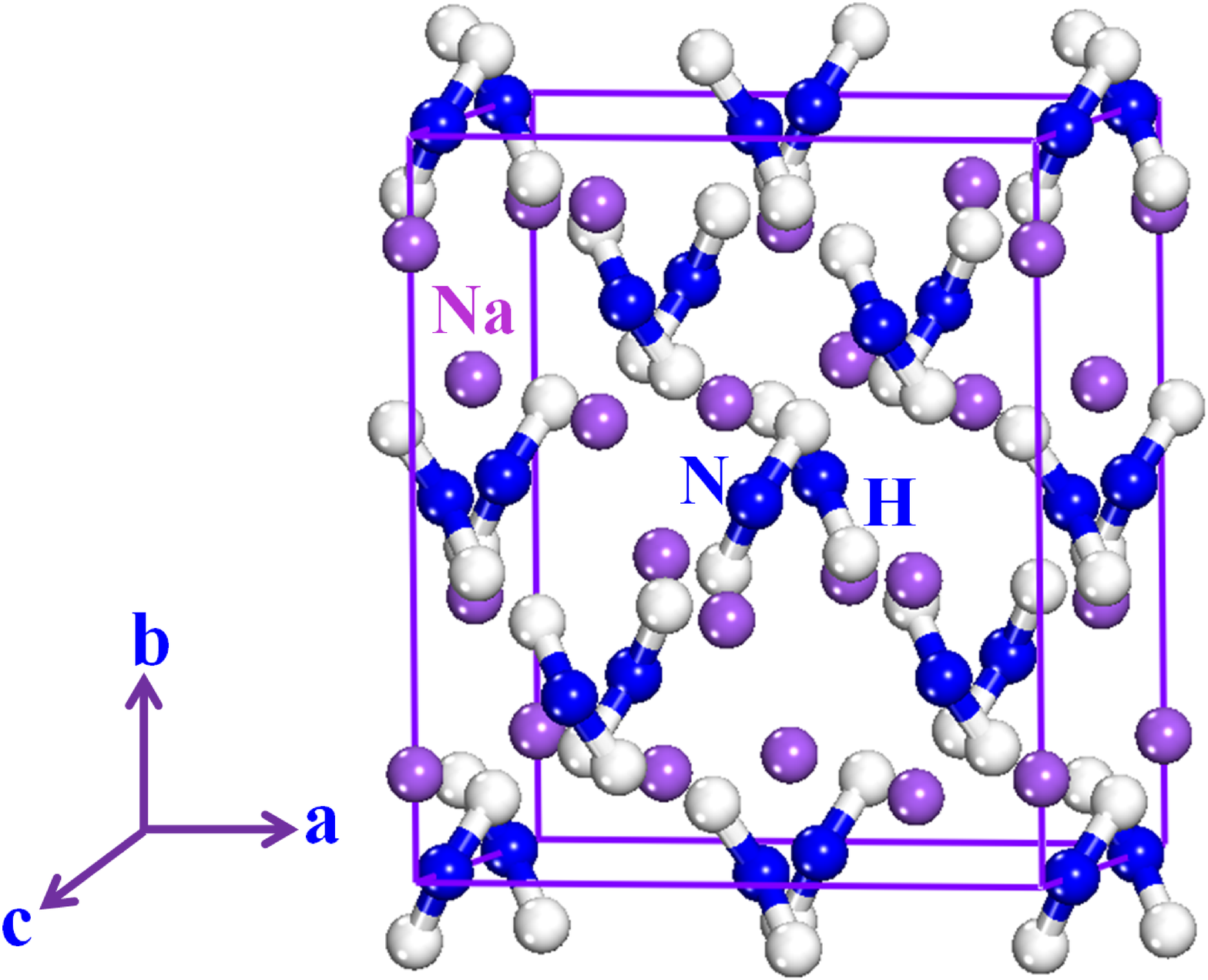}}\\
\subfigure[]{\includegraphics[height=5cm,width=6cm]{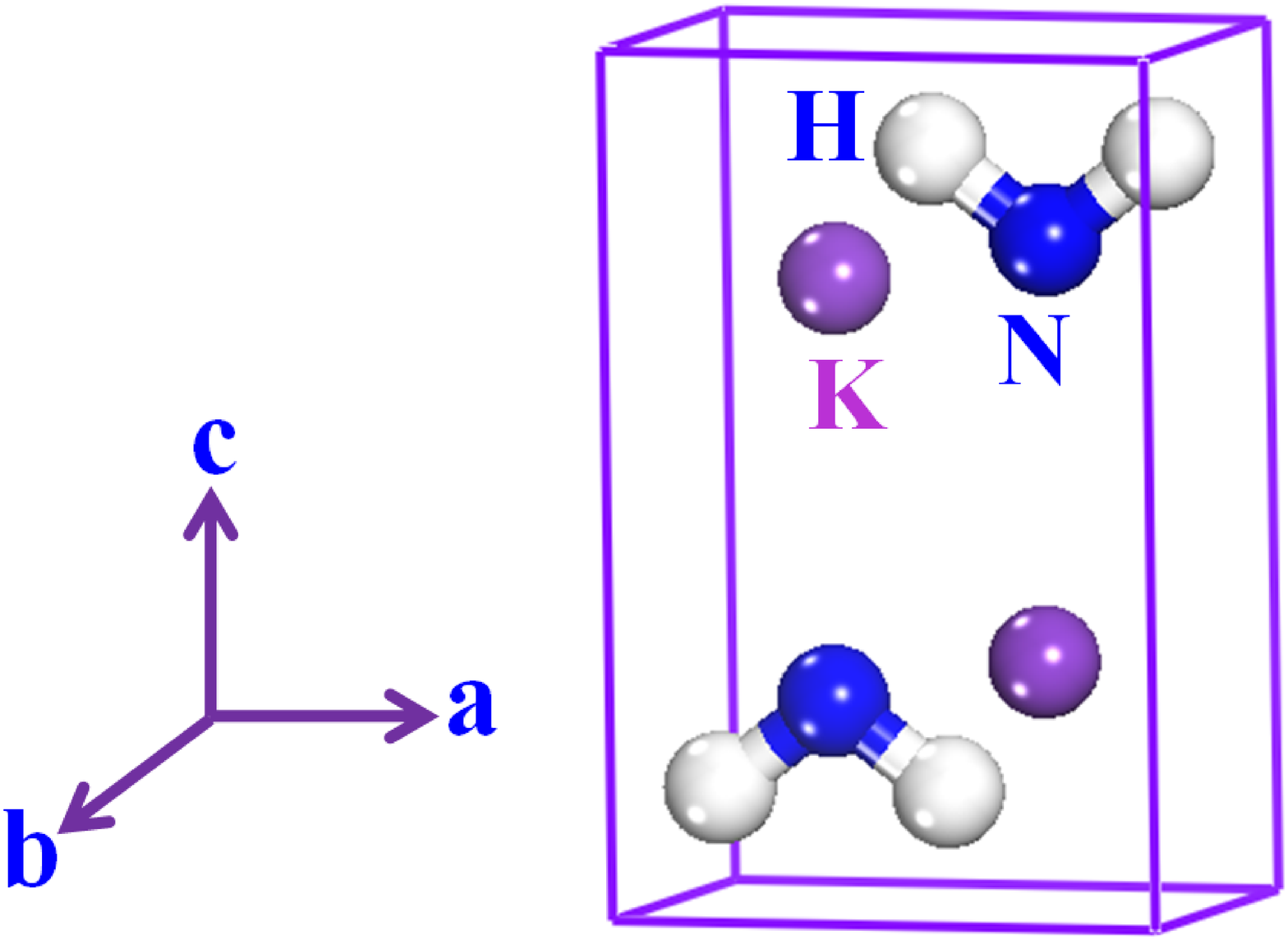}}
 \caption{(Colour online) Crystal Structure of tetragonal LiNH$_2$ (a), orthorhombic NaNH$_2$ (b), and monoclinic KNH$_2$ and RbNH$_2$ (c). In figure, violet ball indicates metal atom, blue ball indicates nitrogen and white ball indicates hydrogen atom. }
 \end{figure}


\begin{table}
\caption{Structural properties of alkali metal amides calculated within LDA (CA-PZ), GGA (PBE) and GGA (PBE+G06).}
\begin{tabular}{ccccccc}\hline \hline
& a (\AA) & b (\AA) & c (\AA)& $\beta$ ($^\circ$) & $\rho$ (g/cc) & V (\AA $^3$)\\ \hline
LiNH$_2$&  & & & & &\\ \hline
LDA (CA-PZ)&4.768 &-- &10.050 &-- &1.330&228.56 \\
GGA (PBE)& 5.181&-- &10.511&-- & 1.081 & 282.25 \\
GGA (PBE+G06)&5.002&--&10.156&--&1.200&254.17 \\
Expt\cite{Jacobs} &5.037 &--&10.278&--&1.169 &260.76 \\ \hline
NaNH$_2$&  & & & & &\\ \hline
LDA (CA-PZ)&8.864 &9.827 &7.438 &-- &1.591 &648.06 \\
GGA (PBE)& 8.920&10.729 &8.450&-- & 1.280 & 808.82 \\
GGA (PBE+G06)&9.103&9.575&7.589&--&1.560&661.63 \\
Expt\cite{Nagib} &8.949 &10.456&8.061&--&1.370 &754.27 \\ \hline
KNH$_2$&  & & & & &\\ \hline
LDA (CA-PZ)&4.317&3.576 &5.910 &95.1 &2.013 &90.92 \\
GGA (PBE)& 4.668&3.805 &6.287&96.3 & 1.649 & 110.99 \\
GGA (PBE+G06)&4.540&3.666&6.017&96.1&1.838&99.60 \\
Expt\cite{Keeffe} &4.586 &3.904&6.223&95.8&1.655 &110.84 \\ \hline
RbNH$_2$&  & & & & &\\ \hline
LDA (CA-PZ)&4.589 &3.775 &6.170 &96.7 &3.174 &106.17 \\
GGA (PBE)& 4.911&4.042&6.573&97.2 & 2.602 & 129.49 \\
GGA (PBE+G06)&4.765&3.812&6.267&97.6&2.985&112.89 \\
Expt\cite{Keeffe} &4.850 &4.148&6.402&97.8&2.641 &127.60 \\ \hline \hline
\end{tabular}
\end{table}

\begin{table}
\caption{Atomic positions of alkali metal amides calculated within LDA (CA-PZ), GGA (PBE) and GGA (PBE+G06).}
\begin{tabular}{ccccc}\hline \hline
& LDA (CA-PZ) & GGA (PBE) & GGA (PBE+G06) & Expt \cite{Jacobs, Nagib, Keeffe} \\ \hline
LiNH$_2$ & & & &  \\ \hline
Li1 & (0 0 0) & (0 0 0)& (0 0 0)& (0 0 0) \\
Li2 & (0 0.5 0.25) & (0 0.5 0.25) & (0 0.5 0.25) & (0 0.5 0.25) \\
Li3 & (0 0.5 0.0010)& (0 0.5 0.0028) & (0 0.5 0.0026) & (0 0.5 0.0042) \\
N & (0.236 0.245 0.117) & (0.2230 0.2494 0.1134) & (0.2366 0.2529 0.1170) & (0.2284 0.2452 0.1148) \\
H1 & (0.230 0.096 0.193) & (0.228 0.129 0.193) & (0.228 0.128 0.198) & (0.226 0.149 0.172) \\
H2 & (0.438 0.327 0.125) & (0.396 0.348 0.120) & (0.428 0.328 0.121) & (0.308 0.359 0.114) \\ \hline
NaNH$_2$ & & & &  \\ \hline
Na & (0 0.1484 0) & (0 0.1400 0)& (0 0.1531 0)& (0 0.1452 0) \\
N & (0 0 0.2509) & (0 0 0.2291) & (0 0 0.2539) & (0 0 0.2365) \\
H & (0.6177 0.7999 0.3677) & (0.6686 0.8347 0.2291) & (0.6031 0.8045 0.3531) & (0.0635 0.9034 0.3053) \\ \hline
KNH$_2$ & & & &  \\ \hline
K & (0.228 0.25 0.315) & (0.222 0.25 0.313)& (0.223 0.25 0.306)& (0.228 0.25 0.295) \\
N & (0.277 0.25 0.770) & (0.271 0.25 0.756) & (0.276 0.25 0.771) & (0.289 0.25 0.778) \\
H & (0.29 0.03 0.88) & (0.31 0.04 0.86) & (0.30 0.03 0.88) & (0.29 0.04 0.88) \\ \hline
RbNH$_2$ & & & &  \\ \hline
Rb & (0.224 0.25 0.310) & (0.214 0.25 0.306)& (0.219 0.25 0.303)& (0.203 0.25 0.295) \\
N & (0.27 0.25 0.77) & (0.27 0.25 0.75) & (0.28 0.25 0.77) & (0.28 0.25 0.79) \\
H & (0.3064 0.0365 0.8859) & (0.3137 0.0516 0.8591) & (0.3195 0.0388 0.8838) & (-- -- --) \\ \hline
\end{tabular}
\end{table}

\begin{figure}
\centering
\subfigure[]{\includegraphics[height=6cm,width=6cm]{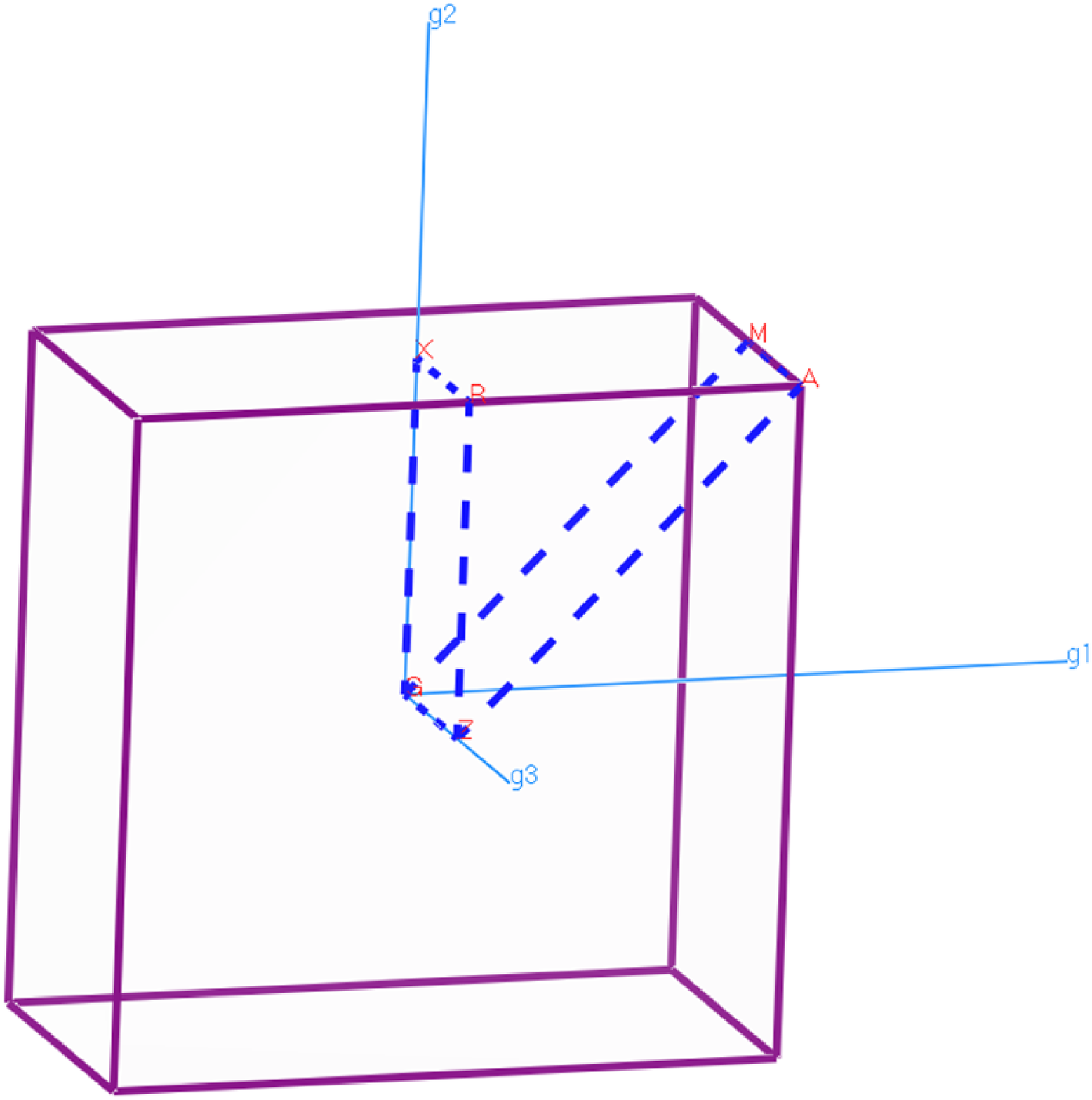}}
\subfigure[]{\includegraphics[height=6cm,width=6cm]{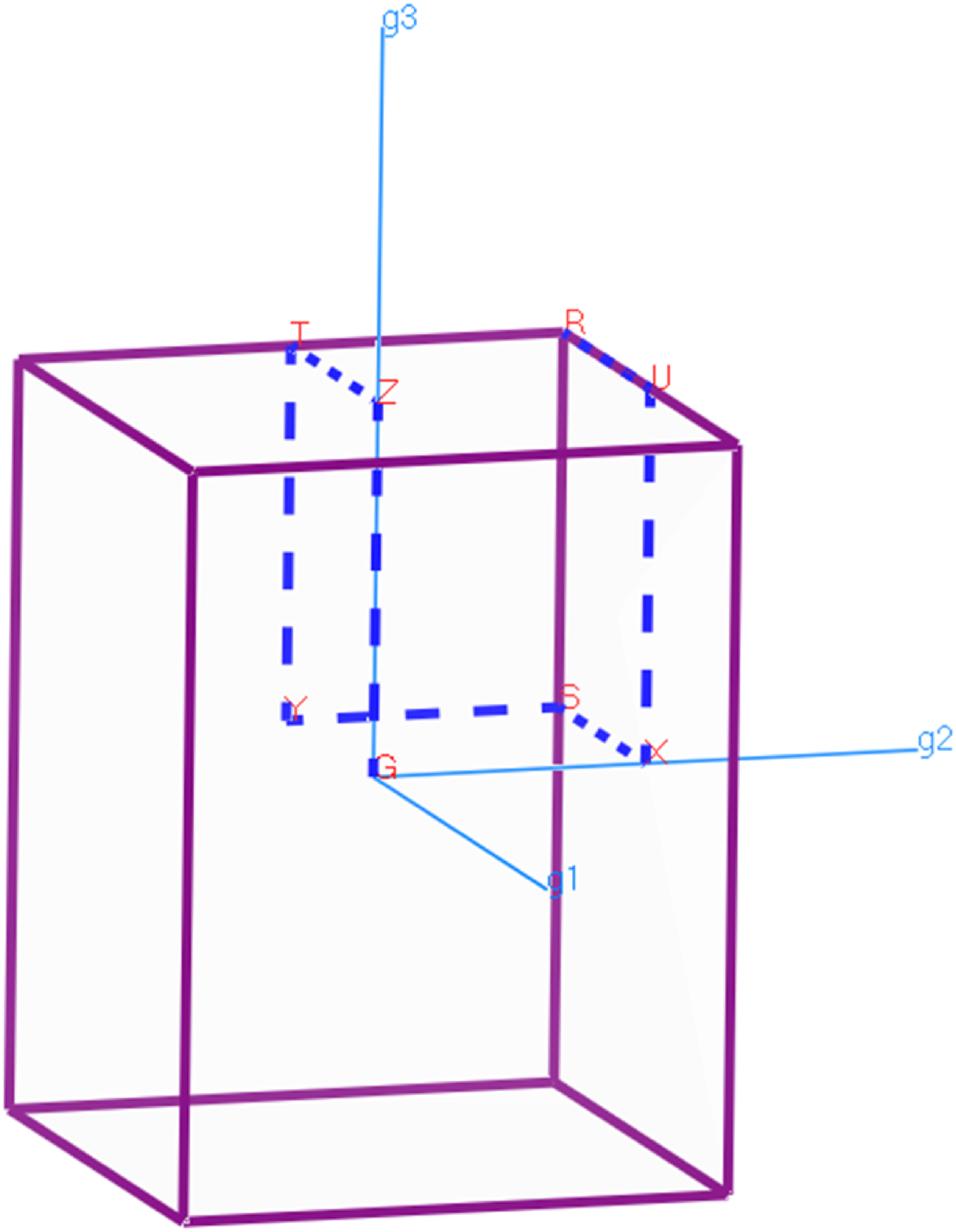}}\\
\subfigure[]{\includegraphics[height=6cm,width=6cm]{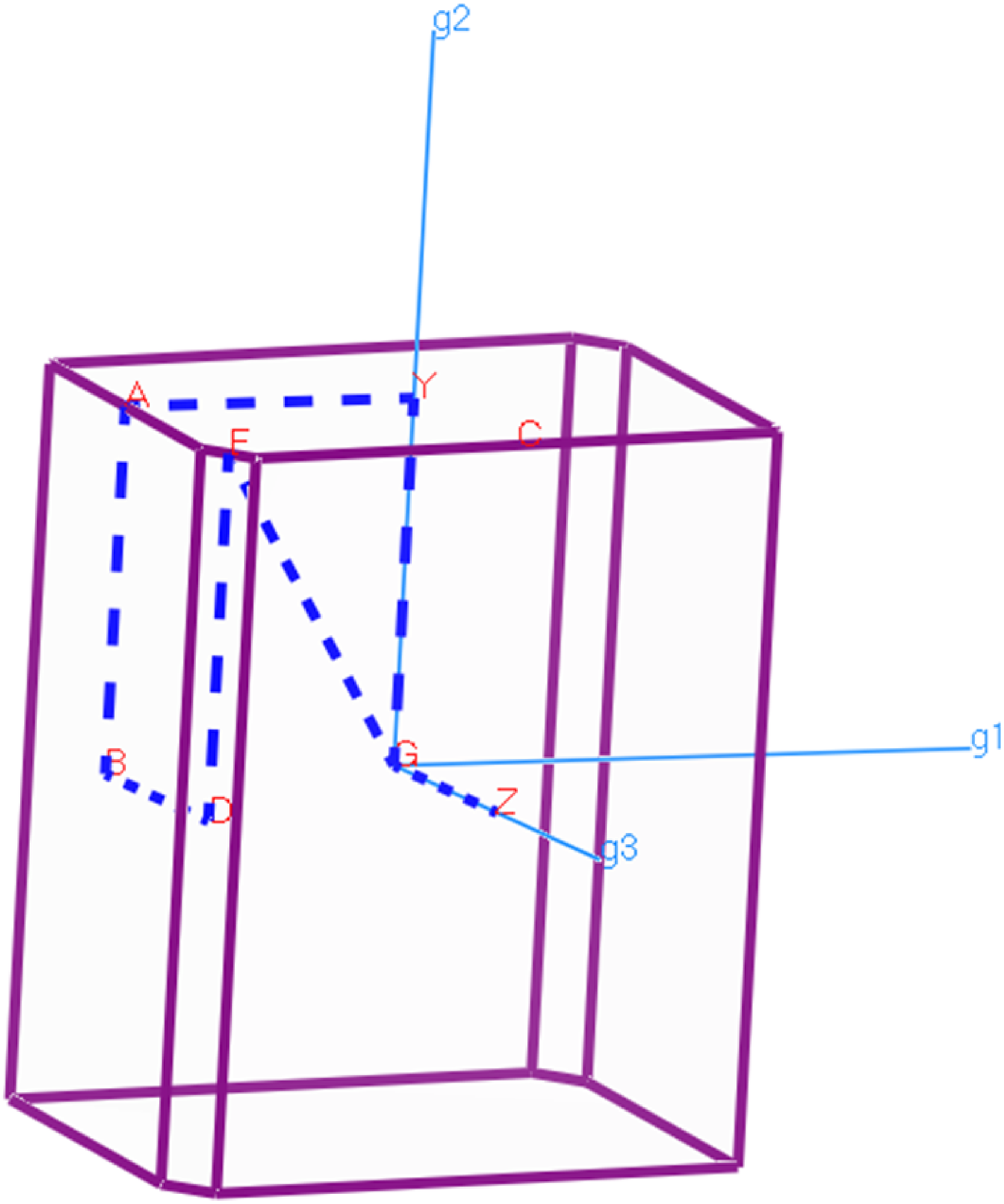}}
 \caption{(Colour online) The reciprocal lattice with Brillouin zone for tetragonal LiNH$_2$ (a), orthorhombic NaNH$_2$ (b), and monoclinic KNH$_2$ and RbNH$_2$ (c). The high symmetry points are connected by blue dotted lines. In figure, g$_1$, g$_2$ and g$_3$ are reciprocal lattice vectors. }
 \end{figure}

\subsection{Elastic properties}
For solid hydrogen storage materials, elastic constants and the related properties are important as they give information regarding the atomic bonding, compressibility characteristics, and phonon properties. Moreover, it was noticed that there is a correlation between the elastic constants and the melting temperature of a solid.  \cite{Nakamori} Hence, it is necessary to get the knowledge of elastic constants of alkali metal amides MNH$_2$ (where `M' is metal atom) for determining their thermodynamic properties, e.g. melting temperature (Tm).
\paragraph*{} By using Hooke's law, for small deformations of a crystal the stress and strain are linearly related by
\begin{equation}
\sigma_{ij} = C_{ijkl} \epsilon_{kl}
\end{equation}
where $\sigma$$_{ij}$ is the stress tensor, $\epsilon$$_{kl}$ is the strain tensor and C$_{ijkl}$ is the elastic stiffness tensor. By following the Voigt's notation\cite{Nye}, this equation can be reduced to
\begin{equation}
\sigma_{x} = C_{xy} \epsilon_{y} \hspace{0.5cm} with \hspace{0.5cm} x=1-6, y= 1-6
\end{equation}
where $xx, yy, zz, yz, xz, xy$ replaced by 1, 2, 3, 4, 5, 6 respectively and therefore C$_{ij}$ form a 6x6 matrix.
 Due to the structural symmetry, the maximum number of independent parameters can be reduced to six for tetragonal structure, nine for orthorhombic structure and thirteen for the monoclinic structure.
\paragraph*{}The calculated elastic constants C$_{ij}$ of MNH$_2$ are displayed in Table III. To the best of our knowledge there are no reports available on the elastic constants of the alkali metal amides except for LiNH$_2$. The computed elastic constants of LiNH$_2$ in the present work are in good comparison with those of earlier theoretical reports based on GGA calculations. \cite{Hector} The knowledge of elastic constants are useful to know about the mechanical stability of the metal amides through Born stability criteria. The calculated C$_{ij}$ satisfy the Born stability criteria for tetragonal, orthorhombic and monoclinic crystals. \cite{Born} Thus, the tetragonal LiNH$_2$, orthorhombic NaNH$_2$ and monoclinic KNH$_2$ and RbNH$_2$ are mechanically stable systems.
\paragraph*{}
For complex hydrides, the elastic constants C$_{11}$, C$_{22}$, and C$_{33}$ are much important as they give necessary information regarding the  atomic bonding characteristics along a, b, and c-axes respectively. For all the compounds we find that the elastic constants follow the order C$_{11}$ $>$ C$_{22}$ $>$ C$_{33}$ except for LiNH$_2$. For the case of LiNH$_2$, the elastic constants C$_{11}$ and C$_{33}$ are almost identical with C$_{33}$ slightly smaller than C$_{11}$  which might be due to the fact that the similar atomic bonding along the (1 0 0) and (0 0 1) planes. Overall, the ordering of elastic constants suggest that the lattice would be more compressible along c-direction when compared to other axes. This fact is confirmed by the recent experimental high pressure study of LiNH$_2$ which suggests that the c-axis is more compressible over a-axis. \cite{Zou} Therefore the present study of elastic constants would be useful for future high pressure experiments on the other metal amides to assess their compressibility behavior. By using the elastic constants, we have calculated the melting temperature (T$_m$) of the amides through the formula given by Fine et al \cite{Fine} T$_m$ = 354 + 4.5 (2C$_{11}$ + C$_{33}$)/3. Clearly, there is a correlation between the bulk modulus B and the melting temperature T$_m$ of the alkali metal amides. The trend of calculated melting temperature T$_m$ of MNH$_2$ follows, RbNH$_2$ (T$_m$ = 464.6 K) $<$ KNH$_2$ (T$_m$ = 481.5 K) $<$ NaNH$_2$ (T$_m$ = 543.8 K) $<$ LiNH$_2$ (T$_m$ = 564.8 K) same as that of bulk modulus B of MNH$_2$ as it follows the order RbNH$_2$ $<$ KNH$_2$ $<$ NaNH$_2$ $<$ LiNH$_2$.


\begin{table}
\caption{The calculated single-crystal elastic constants (SCE) C$_{ij}$ in GPa, polycrystalline elastic moduli (PCE), in GPa of alkali metal amides.}
\begin{tabular}{cccc}\hline \hline
Compd & SCE & PCE \\ \hline
LiNH$_2$ &  C$_{11}$ = 47.4,  C$_{12}$ = 13.3, C$_{13}$ = 17.7, C$_{33}$ = 45.7, \hspace{0.5cm} & B$_H$ = 26.4, G$_H$ = 16.1 \\
 & C$_{44}$ = 14.9,  C$_{66}$ = 20.7\hspace{1cm}  & E = 40.1, B$_H$/G$_H$ = 1.63  \\ \hline
NaNH$_2$ & C$_{11}$ = 53.2, C$_{12}$ = 18.3, C$_{13}$ = 9.6, C$_{22}$ = 20.6, \hspace{0.5cm} & B$_H$ = 14.3, G$_H$ = 6.6 \\
& C$_{23}$ = 2.6, C$_{33}$ = 20.1, C$_{46}$ = 1.6, C$_{55}$ = 10.6, \hspace{0.5cm}  & E = 17.1, B$_H$/G$_H$ = 2.1 \\
& C$_{66}$ = 8.9 \hspace{0.5cm} & \\ \hline
KNH$_2$ & C$_{11}$ = 31.3, C$_{12}$ = 6.5, C$_{13}$ = 6.1, C$_{15}$ = 0.2, \hspace{0.5cm} & B$_H$ = 13.3, G$_H$ = 9.1 \\
& C$_{22}$ = 30.8, C$_{23}$ = 6.2, C$_{25}$ = 0.6, C$_{33}$ = 22.4,  \hspace{0.5cm}  & E = 22.2, B$_H$/G$_H$ = 1.46\\
&  C$_{35}$ = -1.6, C$_{44}$ = 7.1, C$_{46}$ = -0.1, C$_{55}$ = 6.6, \hspace{0.5cm}  & \\
& C$_{66}$ = 11.5  \hspace{1cm} & \\ \hline
RbNH$_2$ & C$_{11}$ = 27.1, C$_{12}$ = 6.1, C$_{13}$ = 7.9, C$_{15}$ = 1.5, \hspace{0.5cm} & B$_H$ = 11.4, G$_H$ = 7.4 \\
& C$_{22}$ = 21.6, C$_{23}$ = 4.4, C$_{25}$ = 1.8, C$_{33}$ = 19.5,  \hspace{0.5cm}  & E = 18.3, B$_H$/G$_H$ = 1.54 \\
&  C$_{35}$ = -1.3, C$_{44}$ = 7.1, C$_{46}$ = -0.4, C$_{55}$ = 5.2, \hspace{0.5cm}  & \\
& C$_{66}$ = 9.7  \hspace{0.5cm} & \\
 \hline \hline
\end{tabular}
\end{table}

\paragraph*{}
The calculated elastic constants allow us to obtain the polycrystalline mechanical properties such as bulk moduli and shear moduli via Voigt-Reuss-Hill approach. \cite{Hill} The calculated polycrystalline elastic properties of MNH$_2$ are presented in Table III. We find that the polycrystalline bulk moduli B$_H$ of the amides are much smaller than typical metals and intermetallic compounds, which indicates that all MNH$_2$ are highly compressible materials. Further, the bulk modulus value decreases with increase in size of the metal atom, implies higher hardness for LiNH$_2$. It is interesting to note that the bulk modulus can be directly correlated with the inverse of the density of the materials. From Table I, the density of the metal amides increases from LiNH$_2$ to RbNH$_2$ which is inverse to the bulk modulus of the compounds as given in Table III. This kind of behaviour was earlier observed in the case of scheelite type AWO$_4$ (A = Ca, Sr, Ba, Pb) compounds. \cite{Javier} Shear moduli G$_H$ represents the strength of the interatomic bonds with respect to the shear deformation, which influences the mobility of dislocations in the solids. The calculated shear modulus of LiNH$_2$ is greater than that of the other metal amides. These results on the bulk moduli and shear moduli of MNH$_2$  follow the trend that was observed in the case of alkali borohydrides. \cite{ZJiang} The calculated B$_H$ and G$_H$ are used to analyze the ductile-brittle nature of the metal amides through Pugh's criterion \cite{Pugh}, according to which the critical value of B$_H$/G$_H$ ratio that separates the ductile and brittle material is 1.75. If B$_H$/G$_H$ $>$ 1.75, the material behaves in a ductile manner, otherwise the material behaves in a brittle manner. For MNH$_2$ (A=Li, K, Rb), the B/G values are less than 1.75 indicating that these complex amides are brittle in nature. However, the B$_H$/G$_H$ value for NaNH$_2$ is greater than 1.75 indicating that the material is ductile. Young's modulus (E) provides a measure of the stiffness of the solid and if the magnitude of E is large, then the material can be regarded as stiffer material. Among the four alkali metal amides, LiNH$_2$ is the stiffest material because of its high value of E. However, when compared to other complex hydrides such as LiBH$_4$ (E=103.68 GPa), LiNH$_2$ has less stiffness.
\subsection{Electronic structure}
The electronic band structures of alkali metal amides MNH$_2$ are calculated by using GGA (PBE) and TB-mBJ functionals. The reason to choose two different functionals is because, in general the energy band gap calculated with standard density functionals such as GGA (PBE) is underestimated by 30 - 40 \% compared to experiment. \cite{Ruiz, Segura} To achieve accurate band gaps that are in good accord with experiment we adopt TB-mBJ functional which is a recently developed semi-local functional and also adjudged as best replacement for computationally expensive GW calculations to predict the band gaps. The computed band structures of LiNH$_2$, NaNH$_2$, KNH$_2$ and RbNH$_2$ along the high symmetry directions in the irreducible Brillouin zone by using the TB-mBJ functional are displayed in Figure 3(a), 3(b), 3(c) and 3(d) respectively. The overall band structure profiles are quite different from each other due to their different crystal geometries. The calculated band structure of LiNH$_2$ shows a band gap of 3.34 eV (PBE) with the maximum of valence band occurs at A-point and minimum of the conduction band occurs at $\Gamma$-point indicating that the compound is an indirect band gap semiconductor. The value of the band gap calculated with PBE is in good agreement with earlier theoretical reports using GGA (3.2 eV). \cite{Miwa} In the case of orthorhombic NaNH$_2$, the calculated band structure shows a gap of 2.09 eV (PBE) between $\Gamma$-$\Gamma$ reveals that the material is a direct gap semiconductor. Both KNH$_2$ and RbNH$_2$ are also found to be indirect gap semiconductors with a gap of 2.07 eV (PBE) and 1.81 eV (PBE) occurs between Z and $\Gamma$ respectively. However, these calculated band gaps with PBE functional are within the limitation of DFT. \cite{TB} The calculated band gaps with TB-mBJ functional are 4.95 eV for LiNH$_2$, 3.55 eV for NaNH$_2$, 3.87 eV for KNH$_2$ and 3.60 eV for RbNH$_2$. Clearly, the band gaps calculated by TB-mBJ functional are improved to a larger extent compared to the PBE band gap values. To the best of our knowledge there are no experimental data available on the band gaps of the materials, we expect that the present study would be of use for future experimental studies.

\begin{figure}
\begin{center}
\subfigure[]{\includegraphics[width=55mm,height=80mm]{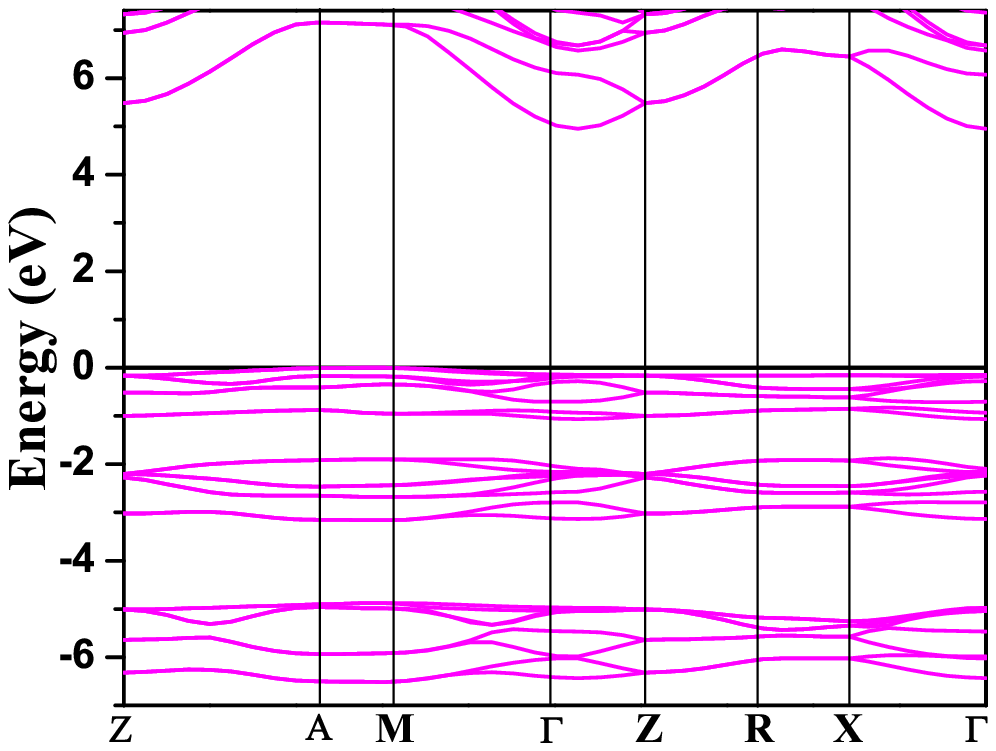}}
\subfigure[]{\includegraphics[width=55mm,height=80mm]{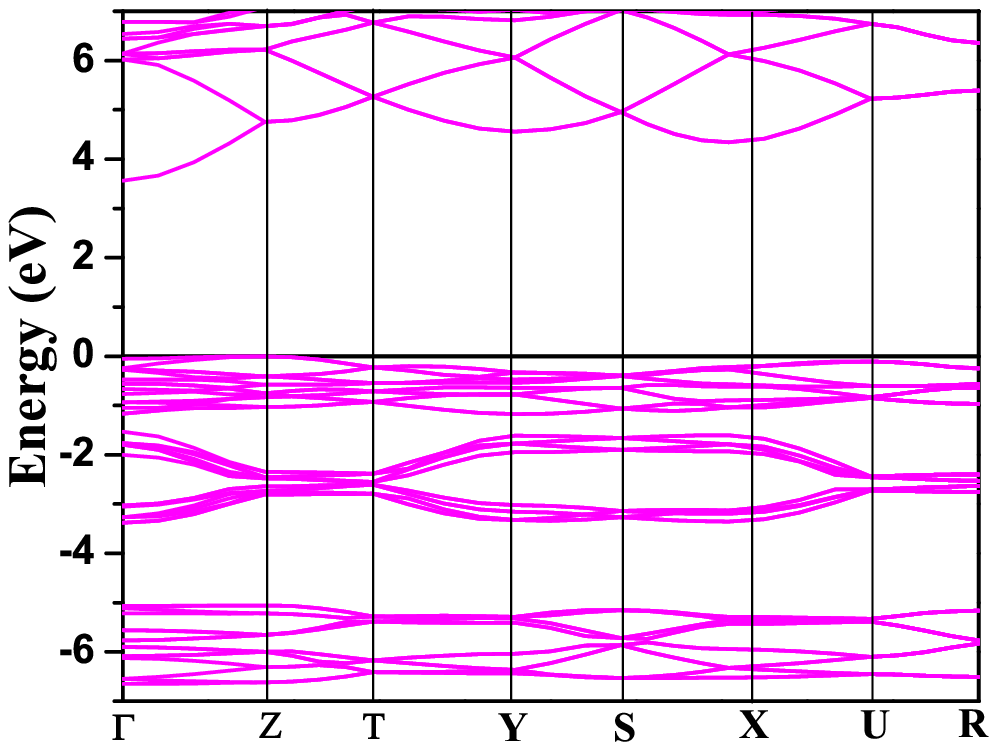}}\\
\subfigure[]{\includegraphics[width=55mm,height=80mm]{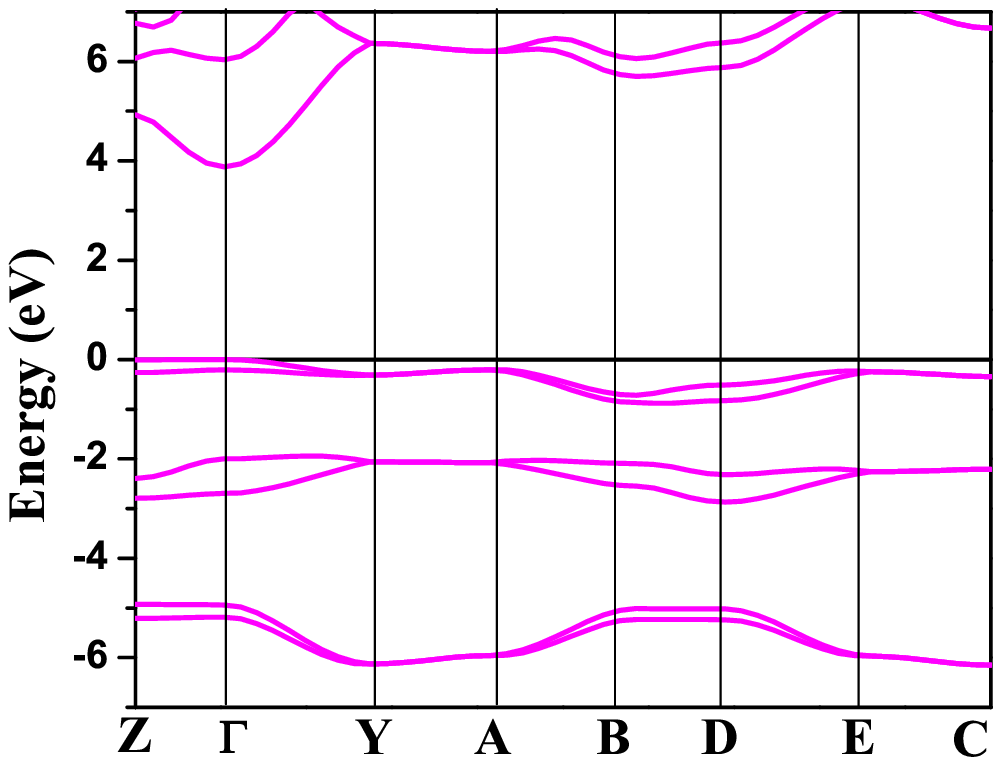}}
\subfigure[]{\includegraphics[width=55mm,height=80mm]{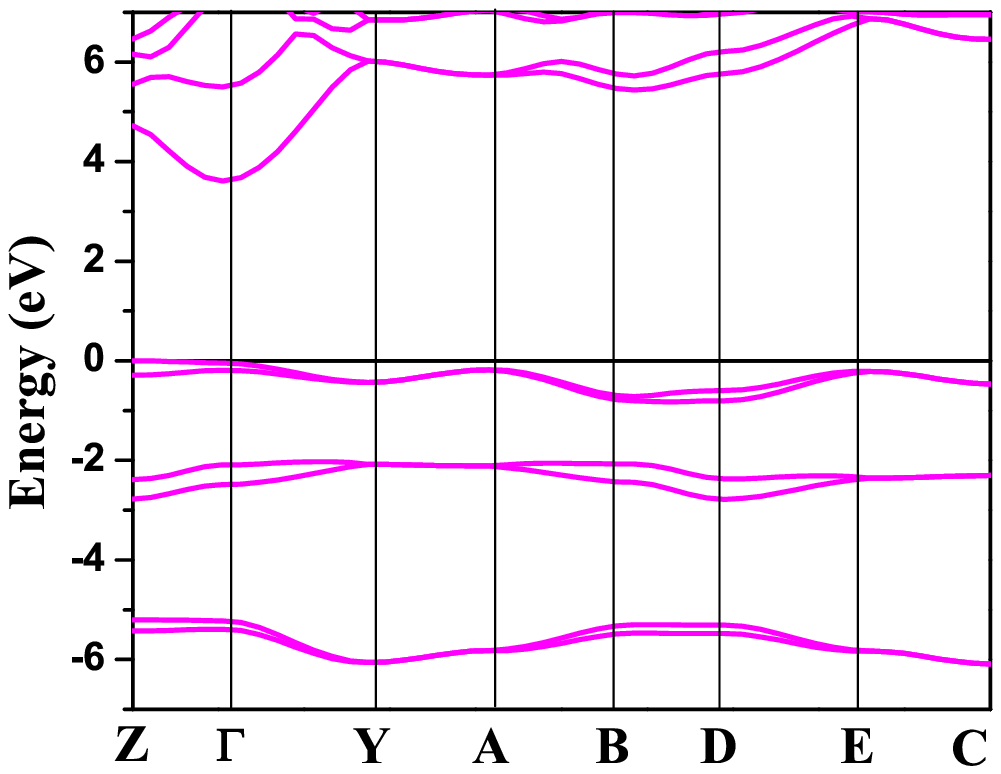}}\\
\caption{(Colour online) Electronic band structure of alkali metal amides LiNH$_2$ (a) NaNH$_2$ (b) KNH$_2$ (c) and RbNH$_2$ (d) calculated at experimental crystal volume.}
\end{center}
\end{figure}

\paragraph*{}
The calculated total and partial density of states of the metal amides MNH$_2$ (M=Li, Na, K, Rb) are shown in Figure 4(a), 4(b), 4(c) and 4(d) respectively. The states near by the Fermi level are mainly dominated by the `p'-states of N atom with a small admixture of `s'-states of metal atom. The lower valence bands that are situated at about -5 to -6 eV are derived from the combination of `s'-states of H atom and `p'-states of N atom. Because of the strong hybridization between the states of H and N, the bands in the valence region are dispersive in nature. In the conduction band the states are entirely from the `s' and `p'- states of metal atom. Overall, it can be noticed that for all the compounds the valence band states are mostly dominated by the states of [NH$_2$]$^-$ anion units, a similar feature also observed in the other complex metal hydrides with a positive cation and anion complex where the hydrogen atom attached covalently with the other group element. \cite{Orimo} A similar phenomena was also observed in the case of alkali metal azides with azide anion units \cite{KRamesh, KRB} which are analogous to alkali metal amides with a positive cation and negative anion.

\begin{figure}
\begin{center}
\subfigure[]{\includegraphics[width=55mm,height=80mm]{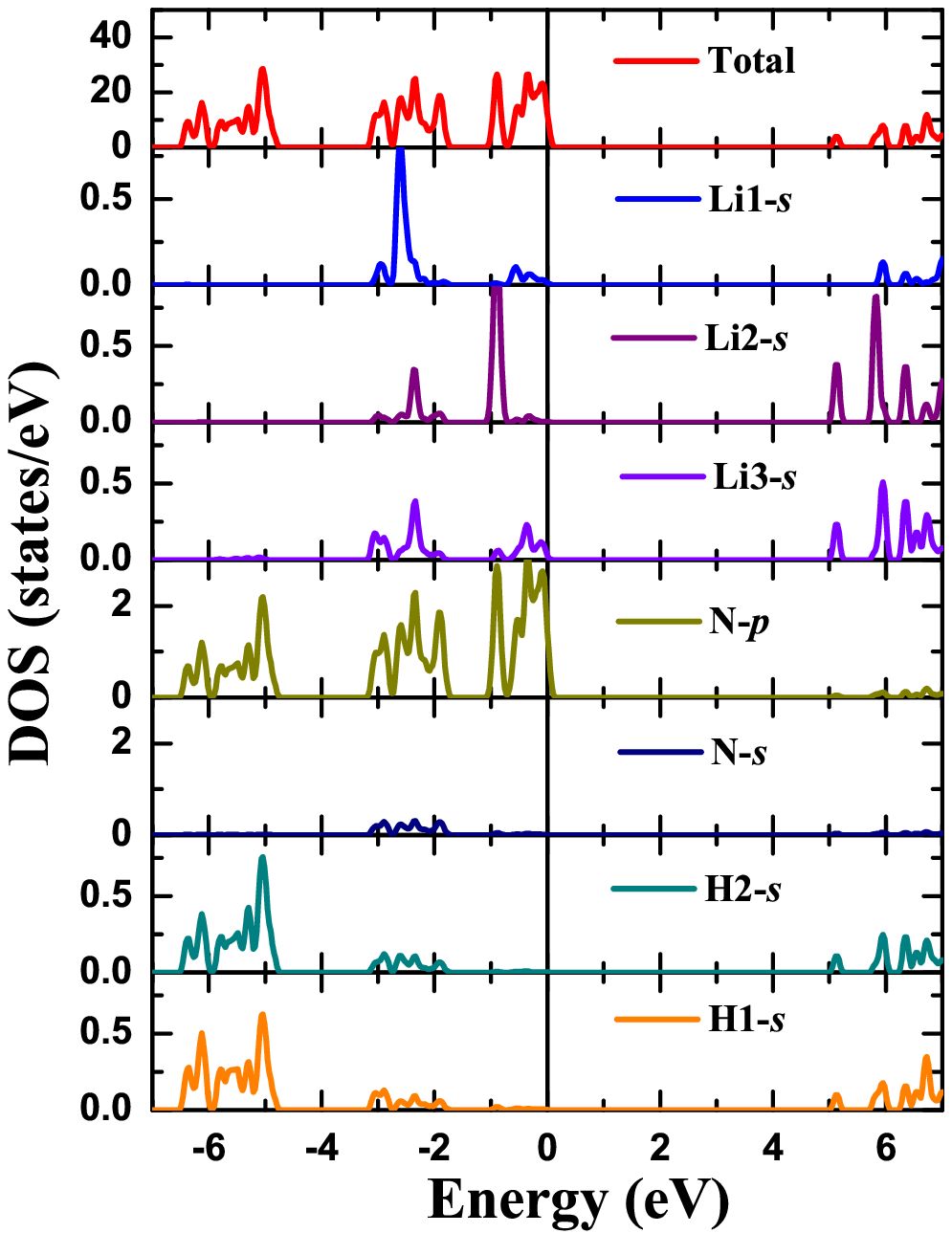}}
\subfigure[]{\includegraphics[width=55mm,height=80mm]{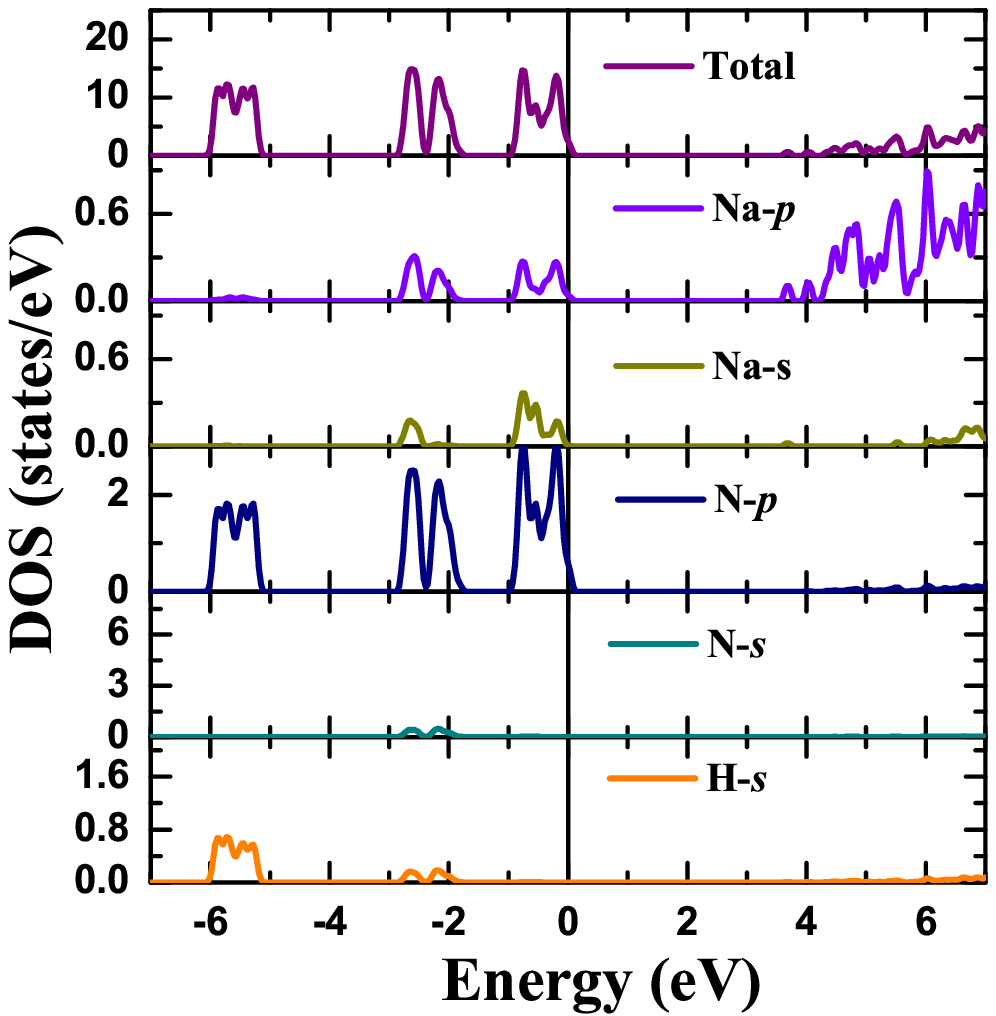}}\\
\subfigure[]{\includegraphics[width=55mm,height=82mm]{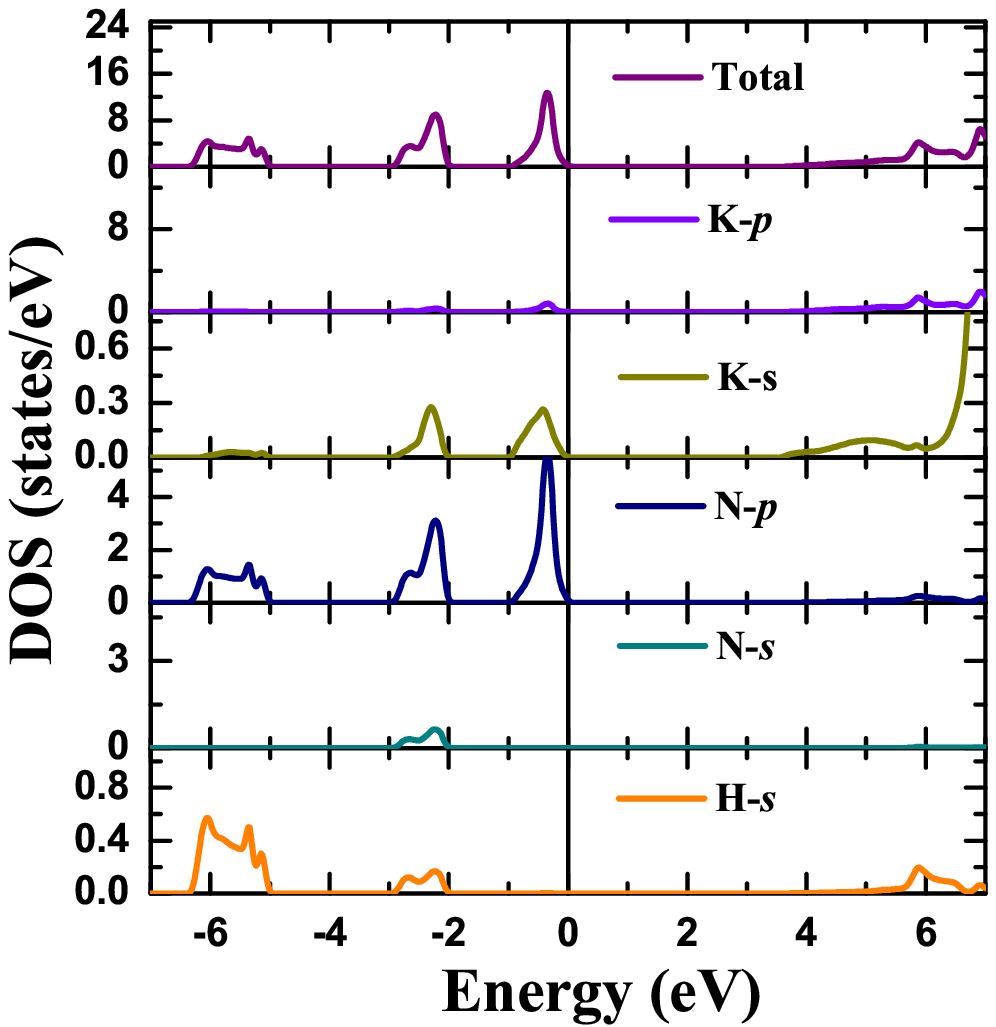}}
\subfigure[]{\includegraphics[width=55mm,height=80mm]{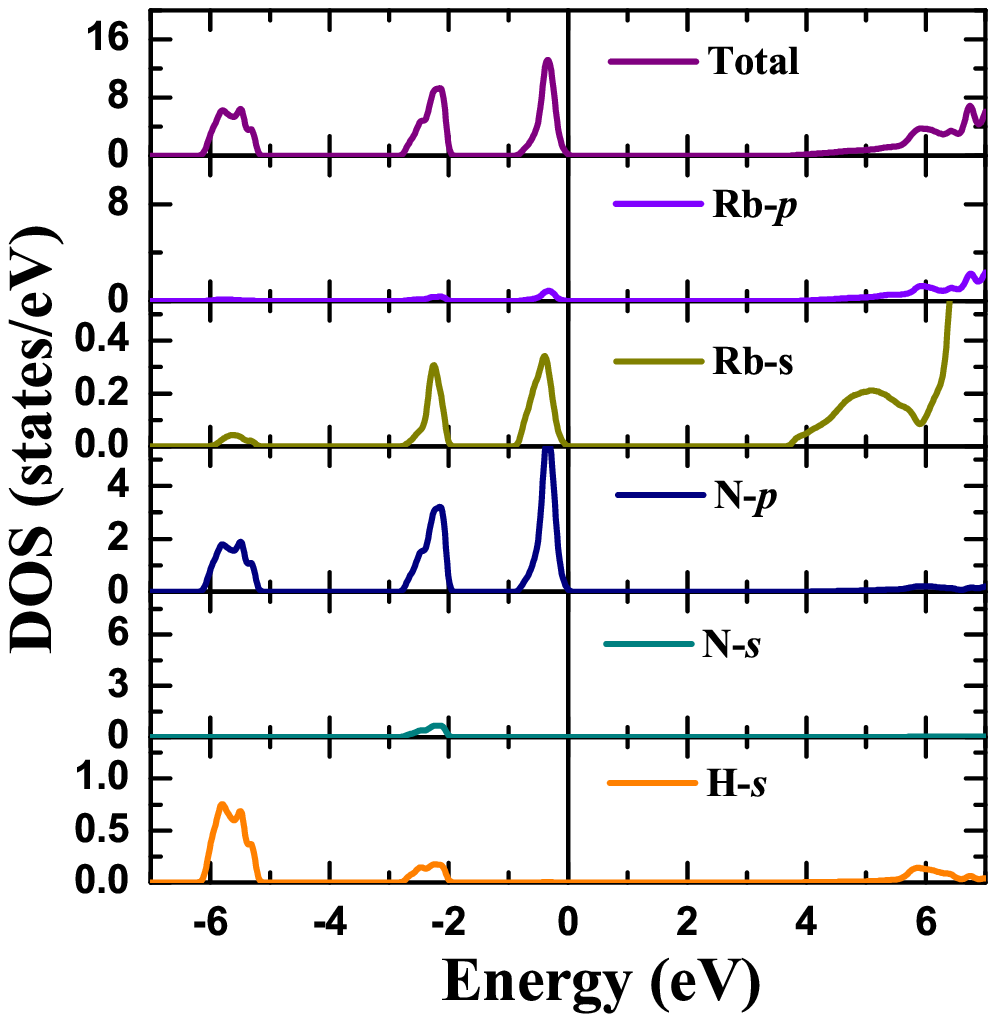}}\\
\caption{(Colour online) Total and partial density of states of alkali metal amides LiNH$_2$ (a) NaNH$_2$ (b) KNH$_2$ (c) and RbNH$_2$ (d).}
\end{center}
\end{figure}

\subsection{Optical properties}
Metal hydrides received tremendous interest by the researchers in recent years due to their potential applications as optical devices. It is thus required to get the knowledge of important optical properties of the metal amide systems. \cite{Koeman} For the metal hydrides LiH, NaH, MgH$_2$, LiAlH$_4$, NaAlH$_4$, and MgAlH$_4$ the optical properties such as complex-dielectric function were reported by using the first principles calculations. \cite{Popa} The optical properties of boron based complex metal hydride NH$_3$BH$_3$ and its metal derivative Ca(NH$_3$BH$_3$)$_2$ were studied by using plane wave pseudopotential method. \cite{Bheem, Tewari} To the best of our knowledge, there are no reports available on the optical properties of the alkali metal amides. Hence, in this present study efforts have been taken to study the optical properties such as complex di-electric function $\epsilon$($\omega$) and refractive index n($\omega$) of alkali metal amides.
\paragraph*{}
In general the optical properties of matter can be described by means of the complex dielectric function $\epsilon (\omega)$ = $\epsilon_1 (\omega)$ + $i\epsilon_2 (\omega)$, where $\epsilon_1 (\omega)$ and $\epsilon_2 (\omega)$ describes the dispersive and absorptive parts of the dielectric function. Normally there are two contributions to $\epsilon(\omega)$ namely intraband and interband transitions. The contribution from intraband transitions is important only for the case of metals. The interband transitions can further be split into direct and indirect transitions. The indirect interband transitions involve scattering of phonons. But the indirect transitions give only a small contribution to $\epsilon(\omega)$ in comparison to the direct transitions, so we neglected them in our calculations. The direct interband contribution to the absorptive or imaginary part $\epsilon_2 (\omega)$  of the dielectric function  $\epsilon(\omega)$ in the random phase approximation without allowance for local field effects can be calculated by summing all the possible transitions from the occupied and unoccupied states with fixed k-vector over the Brillouin zone and is given as \cite{Auluck}
\begin{equation}
\epsilon_2(\omega)=\frac{Ve^2}{2\pi\hbar m^2\omega^2}\int d^3k\sum|\langle\psi_C|p|\psi_V\rangle|^2\delta(E_C-E_V-\hbar\omega)
\end{equation}
 here  $\psi_C$ and  $\psi_V$ are the wave functions in the conduction and valence bands, $p$ is the momentum operator, $\omega$  is the photon frequency, and $\hbar$  is reduced Planck's constant.
The real part $\epsilon_1$ ($\omega$) of the dielectric function can be evaluated from $\epsilon_2$ ($\omega$) by using the Kramer-Kroning relations.
\begin{equation}
\epsilon_1(\omega)=1+\frac{2}{\pi}P\int_0^\infty \frac{\epsilon_2(\omega^\prime)\omega^\prime d\omega^\prime}{(\omega^\prime)^2-(\omega)^2}
\end{equation}
where `P' is the principle value of the integral. The knowledge of both the real and imaginary parts of the dielectric function allows the calculation of the important optical properties such as refractive index $n$($\omega)$ through the following equation
\begin{equation}
 n(\omega) = \frac{1}{\sqrt{2}}\bigg[\sqrt{\epsilon_1(\omega)^2+\epsilon_2(\omega)^2}+\epsilon_1(\omega)\bigg]^\frac{1}{2}
\end{equation}
\paragraph*{}
To calculate the optical properties, it is important to use a sufficient number of k-points in the Brillouin zone because the matrix element changes more rapidly within the Brillouin zone than electronic energies themselves. Therefore, one requires more k-points to integrate this property accurately than are needed for an ordinary SCF calculation. Hence in this study we use a k-point mesh of 8x8x6, 8x7x8, 12x12x12 for LiNH$_2$, NaNH$_2$, and KNH$_2$, RbNH$_2$ respectively for the optical properties calculation. The calculated optical spectra using PBE functional have shifted by scissors operator with a magnitude of difference of TB-mBJ and PBE band gap values. In Figure 5, we have shown the calculated imaginary part of the $\epsilon$($\omega$) as a function of photon energy up to 30 eV. From Figure 5, it can be seen that the threshold of $\epsilon_2 (\omega)$ occurring at the band gap value of the material and this corresponds to the interband optical transition between N-`p' states occupied at the top of the valence band to the lowest occupied conduction band states of M-`s'. In all the four compounds the highest peak in $\epsilon_2 (\omega)$ occurring at around 10 eV is due to the optical transition from the occupied H-`s' states in valence band to the unoccupied M-`p' states in conduction band. The calculated dependence of refractive index of the metal amides on the photon energy is displayed in Figure 6. The refractive index along the three polarizations follows that n$_{[100]}$(0) $\neq$ n$_{[010]}$(0) $\neq$  n$_{[001]}$(0) for NaNH$_2$, KNH$_2$, RbNH$_2$  and n$_{[100]}$(0) $\neq$ n$_{[001]}$(0) for LiNH$_2$, which implies that the materials are optically anisotropic. For all the compounds the refractive index reaches maximum in near Ultra-Violet region (5 to 10 eV). When the refractive index increases with photon energy the compounds show anomalous dispersion and this is find in the visible and near Ultra-Violet region of the spectrum. The characteristic of the anomalous dispersion of the materials is due to the interband optical absorption.

\begin{figure}
\begin{center}
\subfigure[]{\includegraphics[width=80mm,height=80mm]{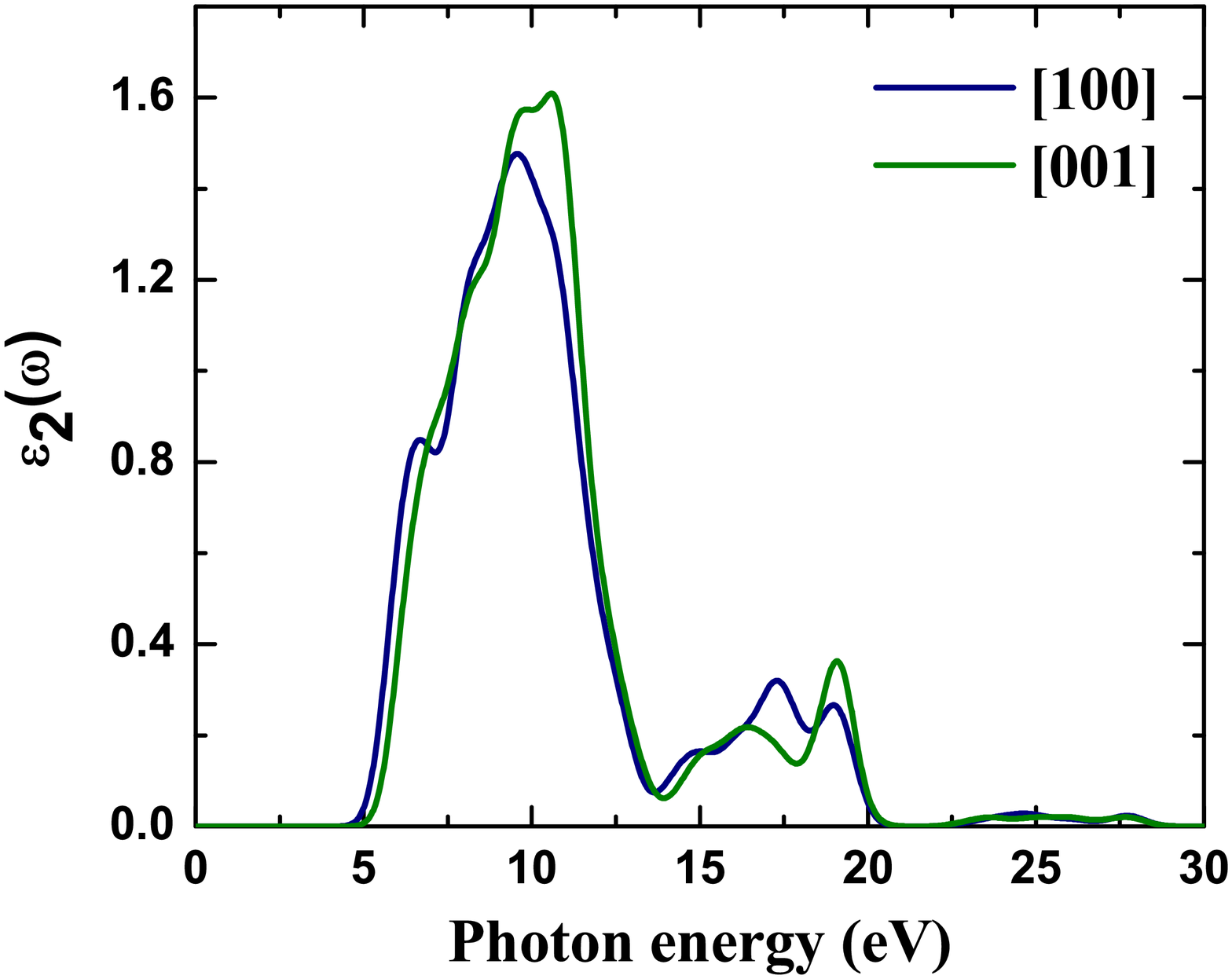}}
\subfigure[]{\includegraphics[width=80mm,height=80mm]{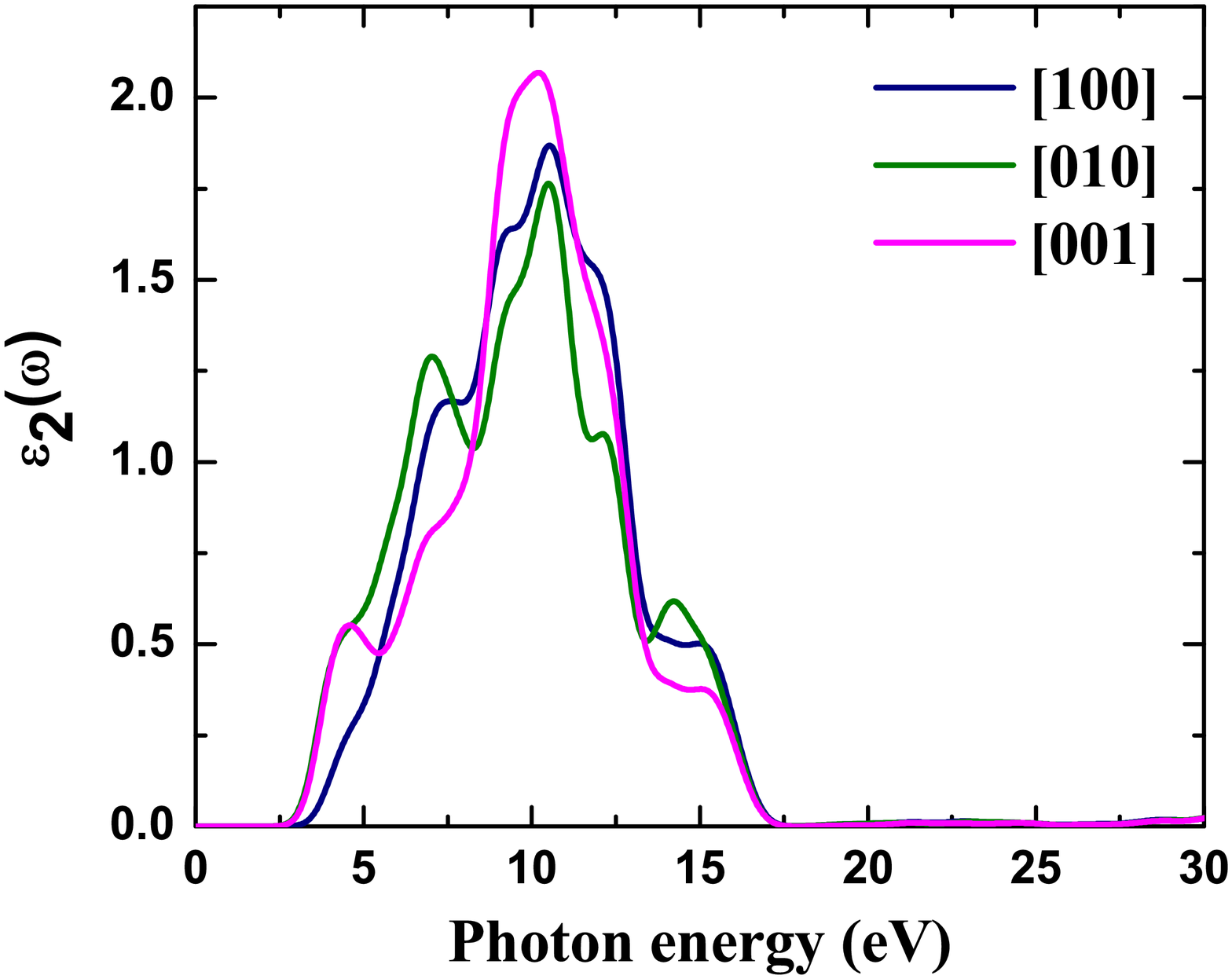}}\\
\subfigure[]{\includegraphics[width=80mm,height=80mm]{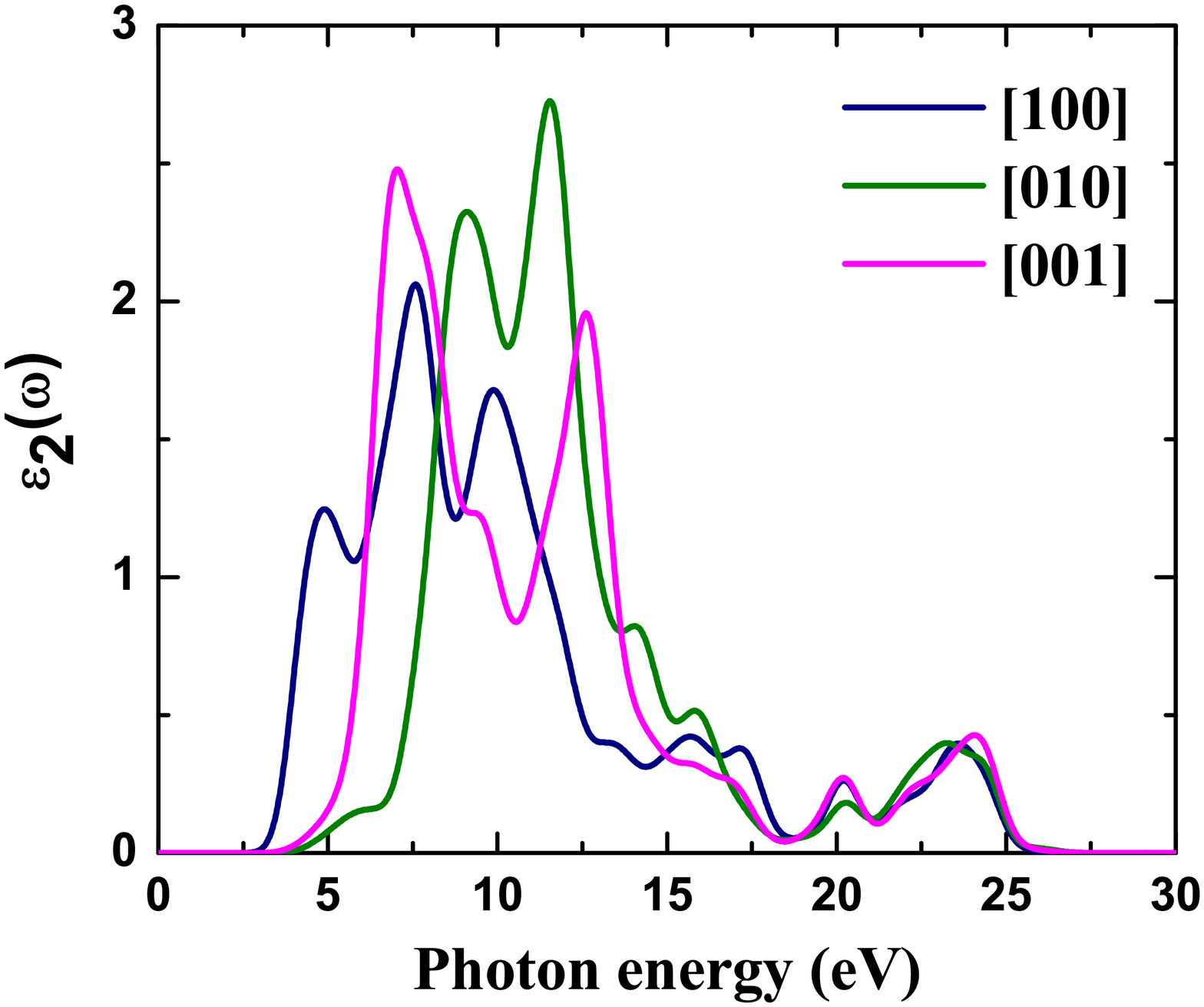}}
\subfigure[]{\includegraphics[width=80mm,height=80mm]{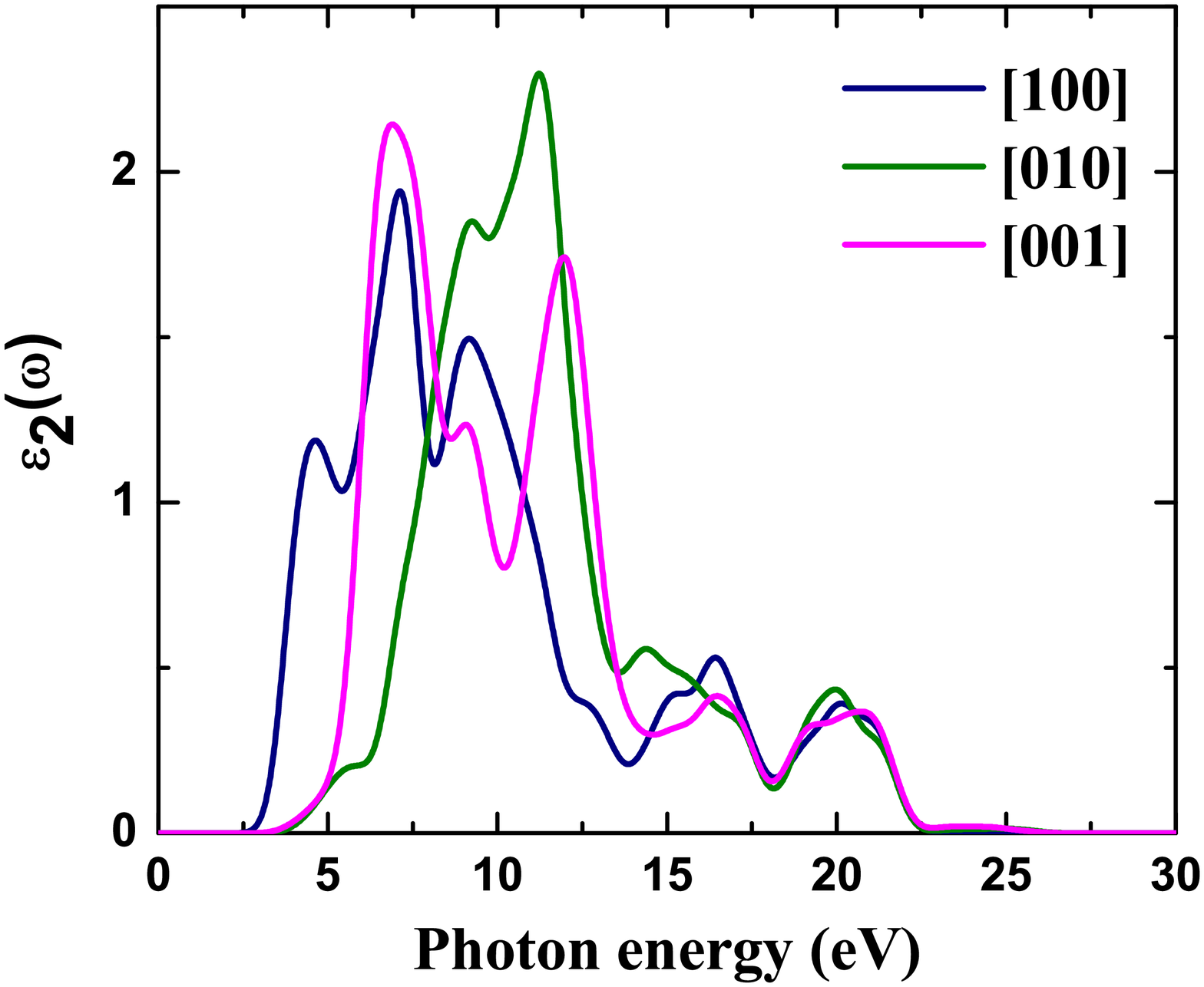}}\\
\caption{(Colour online) The imaginary part of the dielectric function $\epsilon$$_2$($\omega$) of alkali metal amides LiNH$_2$ (a), NaNH$_2$ (b), KNH$_2$ (c) and RbNH$_2$ (d) as a function of photon energy calculated within the PBE functional with a scissors shift at experimental volume.}
\end{center}
\end{figure}

\clearpage
\newpage

\begin{figure}
\begin{center}
\subfigure[]{\includegraphics[width=80mm,height=80mm]{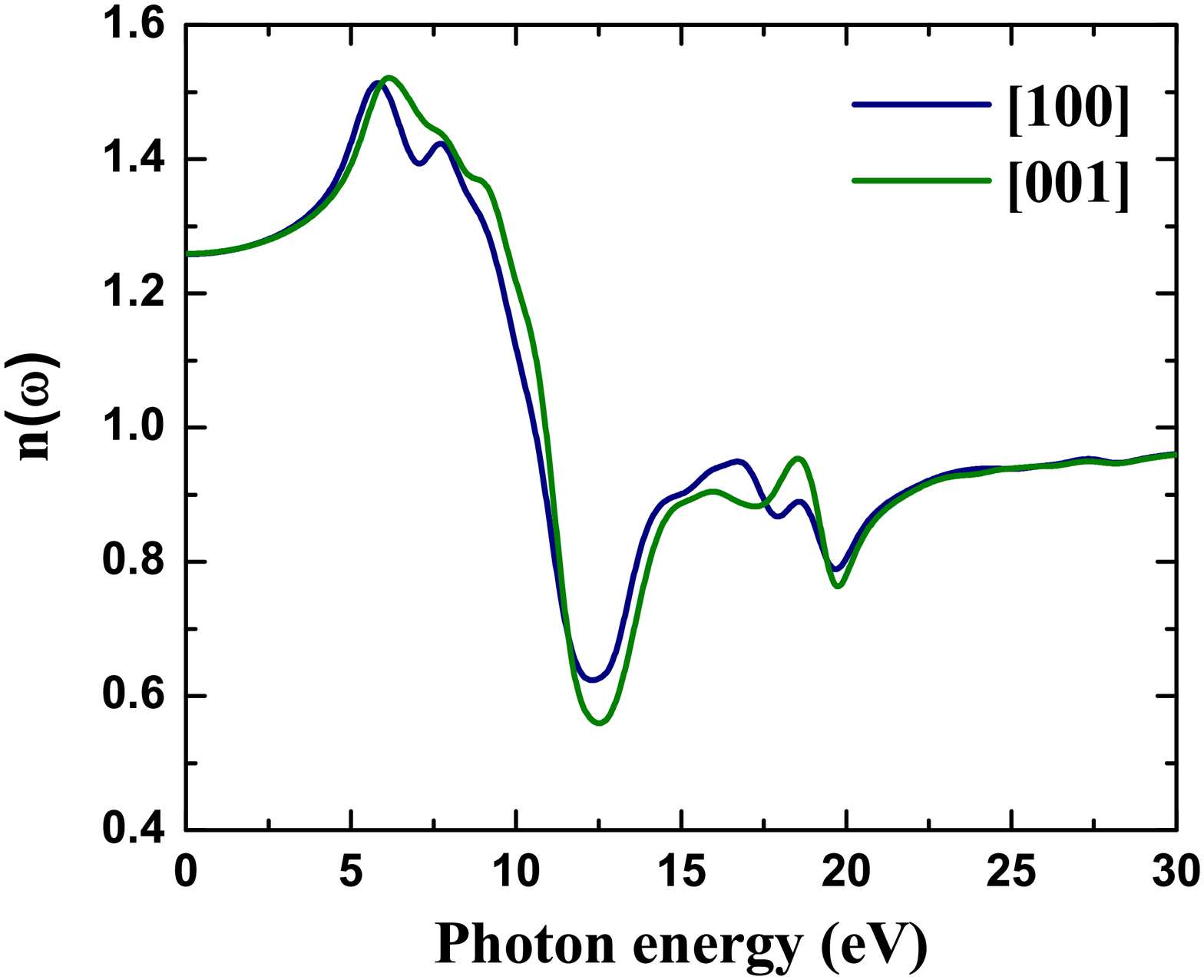}}
\subfigure[]{\includegraphics[width=80mm,height=80mm]{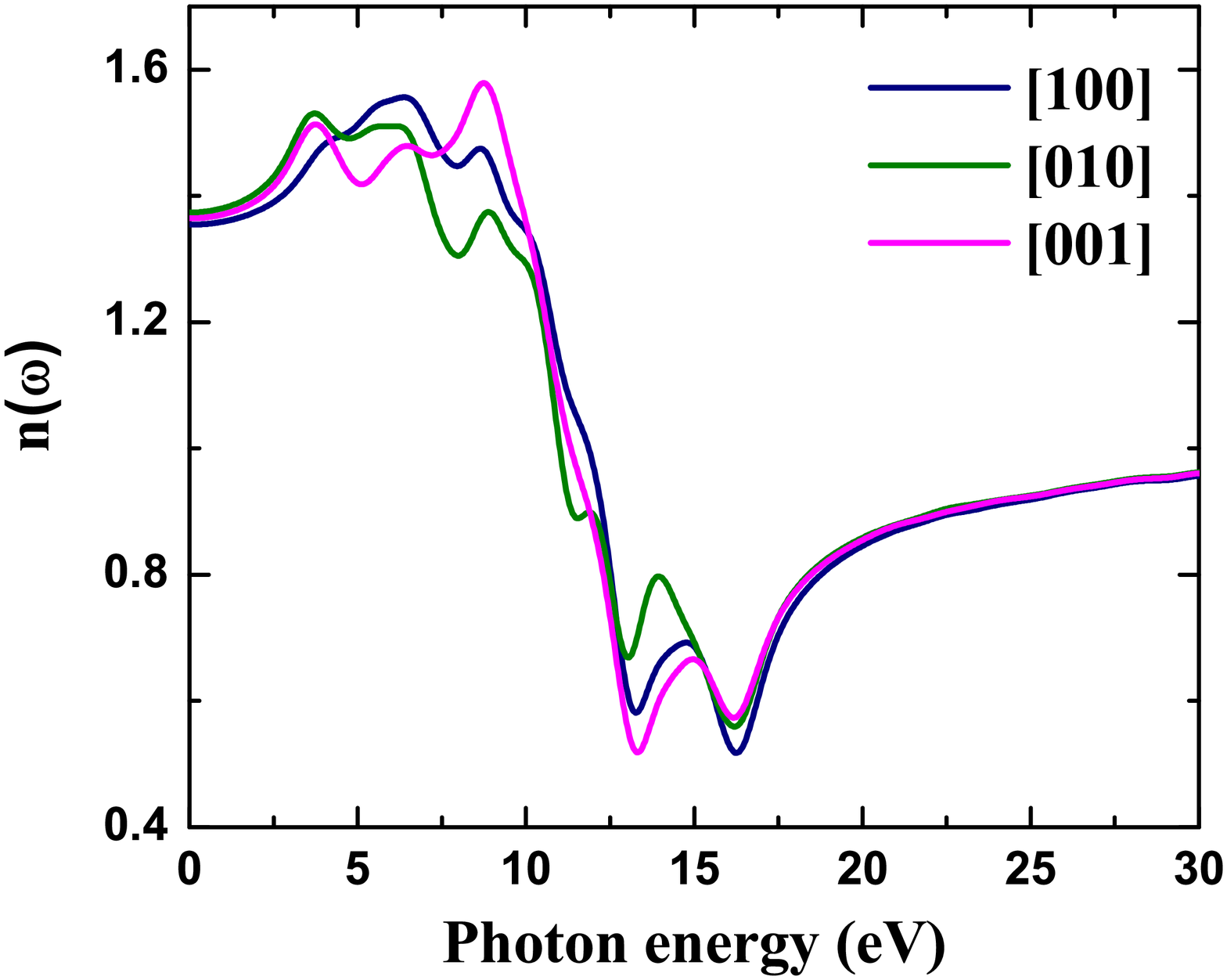}}\\
\subfigure[]{\includegraphics[width=80mm,height=80mm]{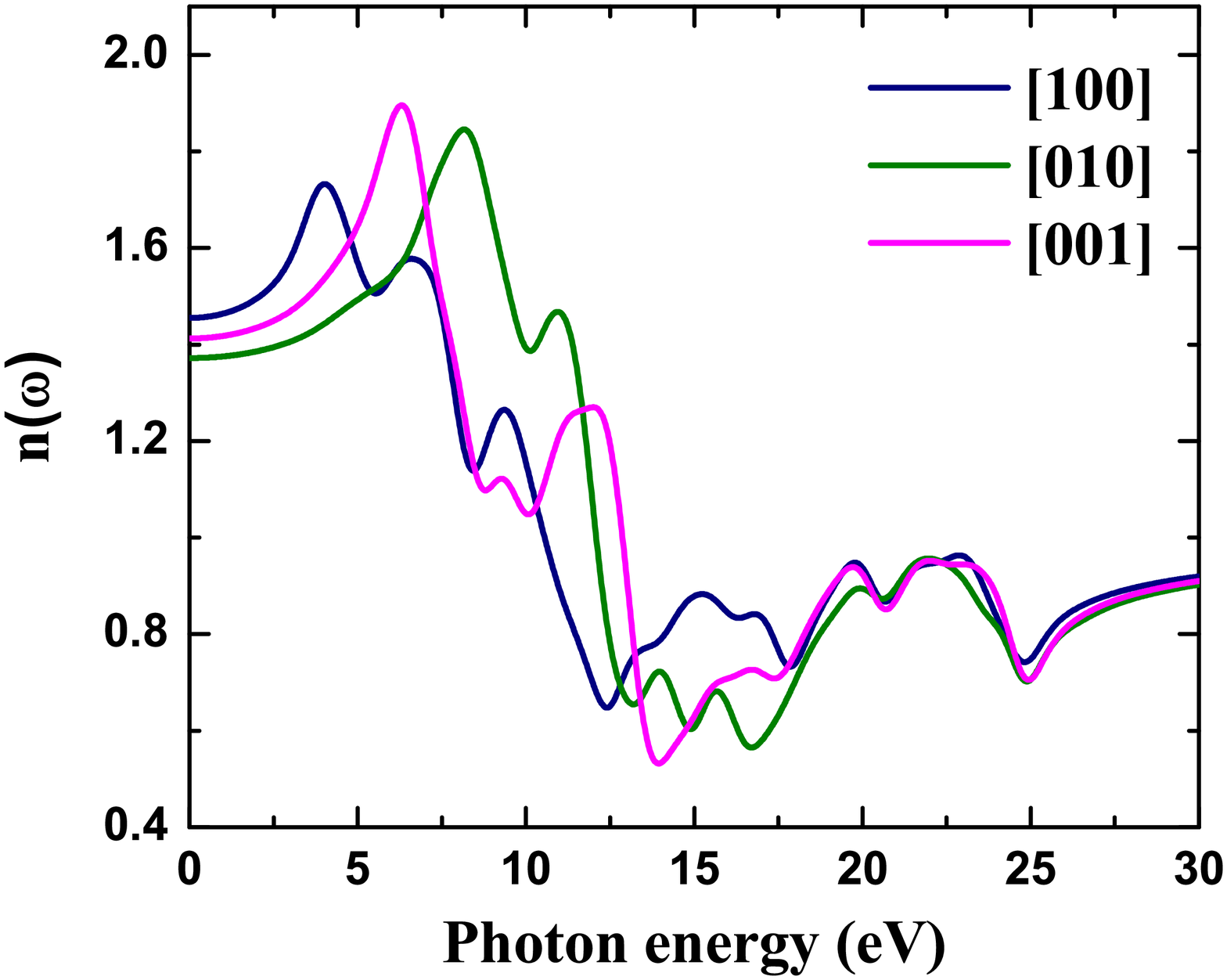}}
\subfigure[]{\includegraphics[width=80mm,height=80mm]{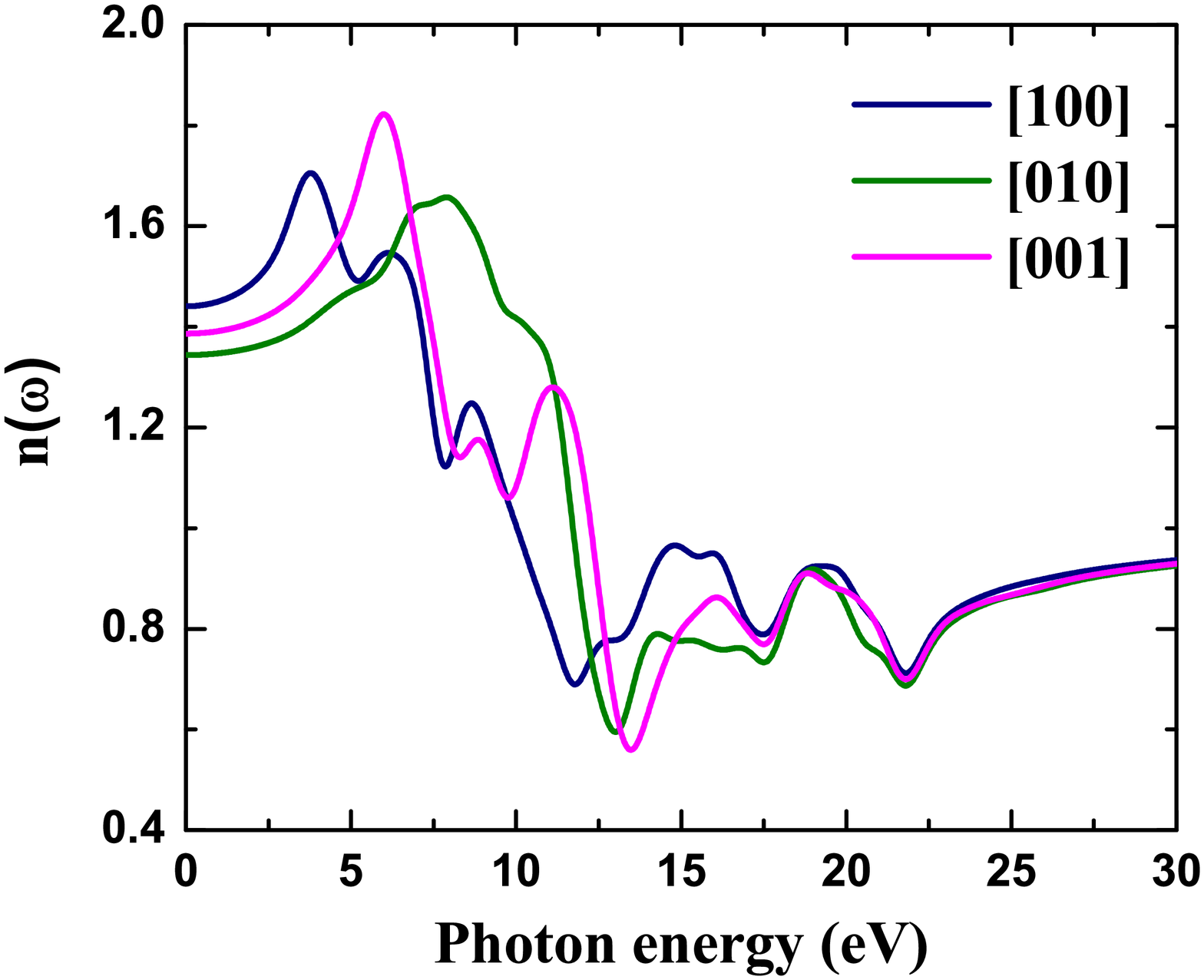}}\\
\caption{(Colour online) Refractive index n($\omega$) of alkali metal amides LiNH$_2$ (a), NaNH$_2$ (b), KNH$_2$ (c) and RbNH$_2$ (d).as a function of photon energy calculated within the PBE functional with a scissors shift at experimental volume.}
\end{center}
\end{figure}

\paragraph*{}

\section{Conclusions}
In conclusion, we have studied the structural, electronic, elastic and optical properties of alkali metal amides. We find that GGA functional with PBE parameterizations give better results when compared to the LDA functional. Among the alkali metal amides LiNH$_2$ has larger bulk modulus due to which the material is stiffer when compared to the other amides. From the calculation of elastic constants, we conclude that all the materials are mechanically stable. Most importantly, the lattice are stiffer along a-axis over the other axis of the metal amides. The melting temperatures, Tm are calculated through the elastic constants and found that Tm follows the order of bulk modlui of the compounds. The electronic band structures are calculated by means of PBE and TB-mBJ functional and found that all the studied materials have nonconducting nature. The valence region of the band structure is entirely dominated by the states of [NH$_2$]$^-$ complex with $p$-states of nitrogen atoms at the Fermi level. The optical properties of the amides are calculated and analyzed. Our study show that the compounds show considerable optical anisotropy in the studied properties.

\section{ACKNOWLEDGMENTS}
 K R B would like to thank DRDO through ACRHEM for financial support. The authors acknowledge CMSD, University of Hyderabad for providing computational facilities.
 \clearpage
*Author for Correspondence,
E-mail: gvsp@uohyd.ernet.in

\end{document}